\documentclass[fleqn,usenatbib]{mnras}

\usepackage{newtxtext,newtxmath}

\usepackage[T1]{fontenc}

\DeclareRobustCommand{\VAN}[3]{#2}
\let\VANthebibliography\thebibliography
\def\thebibliography{\DeclareRobustCommand{\VAN}[3]{##3}\VANthebibliography}

\usepackage{graphicx}	
\usepackage{amsmath}	
\usepackage{xcolor,color,soul,soulutf8}
\sethlcolor{yellow}

\title[The exoplanet mass-radius relationship]{Revisiting mass-radius relationships for exoplanet populations: a machine learning insight} 

\author[M.~Mousavi-Sadr et al.]{
M.~Mousavi-Sadr,$^{1}$\thanks{E-mail: mahdiyar.mousavi@gmail.com}
D.~M.~Jassur,$^{1}$
and G.~Gozaliasl$^{2, 3}$\thanks{E-mail: ghassem.gozaliasl@helsinki.fi}
\\
$^{1}$Department of Theoretical Physics and Astrophysics, Faculty of Physics, University of Tabriz, Tabriz, Iran\\
$^{2}$Department of Computer Science, Aalto University, P. O. Box 15400, Espoo, FI-00076, Finland\\
$^{3}$Department of Physics, University of Helsinki, P. O. Box 64, FI-00014, Helsinki, Finland\\
}

\date{Accepted XXX. Received YYY; in original form ZZZ}

\pubyear{2015}

\begin{document}

\defcitealias{2019A&A...630A.135U}{Ulmer19}

\label{firstpage}
\pagerange{\pageref{firstpage}--\pageref{lastpage}}
\maketitle

\begin{abstract}
The growing number of exoplanet discoveries and advances in machine learning techniques have opened new avenues for exploring and understanding the characteristics of worlds beyond our Solar System. In this study, we employ efficient machine learning approaches to analyze a dataset comprising 762 confirmed exoplanets and eight Solar System planets, aiming to characterize their fundamental quantities. By applying different unsupervised clustering algorithms, we classify the data into two main classes: “small” and “giant” planets, with cut-off values at $R_{p}=8.13R_{\oplus}$ and $M_{p}=52.48M_{\oplus}$. This classification reveals an intriguing distinction: giant planets have lower densities, suggesting higher H-He mass fractions, while small planets are denser, composed mainly of heavier elements. We apply various regression models to uncover correlations between physical parameters and their predictive power for exoplanet radius. Our analysis highlights that planetary mass, orbital period, and stellar mass play crucial roles in predicting exoplanet radius. Among the models evaluated, the Support Vector Regression consistently outperforms others, demonstrating its promise for obtaining accurate planetary radius estimates. Furthermore, we derive parametric equations using the M5P and Markov Chain Monte Carlo methods. Notably, our study reveals a noteworthy result: small planets exhibit a positive linear mass-radius relation, aligning with previous findings. Conversely, for giant planets, we observe a strong correlation between planetary radius and the mass of their host stars, which might provide intriguing insights into the relationship between giant planet formation and stellar characteristics.
\end{abstract}



\begin{figure}
	\includegraphics[width=1.0\columnwidth]{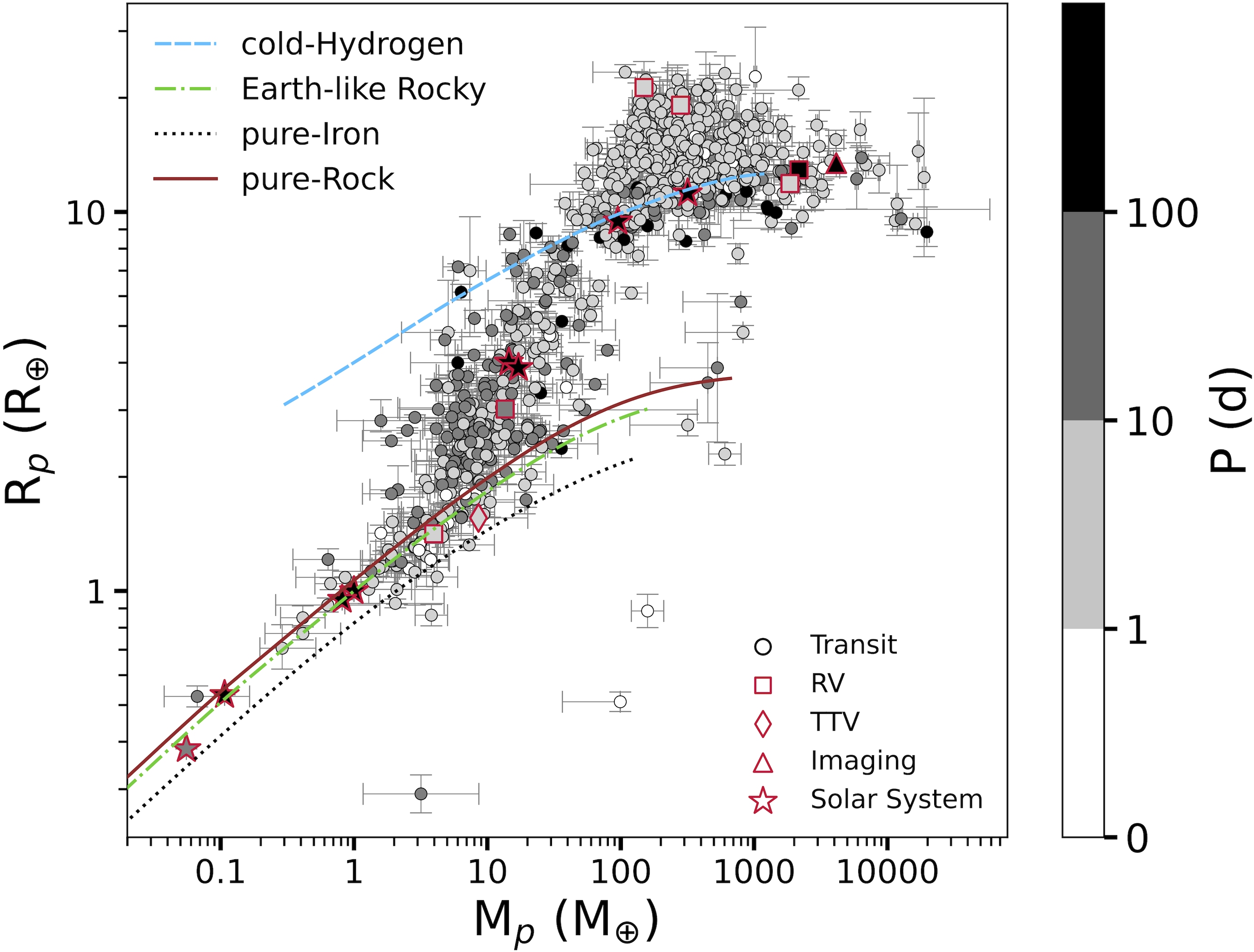}
    \caption{The mass-radius distribution of 770 planets color coded by orbital period. This figure separates exoplanets into four groups based on detection methods: transit (black circles), radial velocity (red squares), transit timing variations (red diamond), and imaging (red triangle). Red stars show the Solar System's planets. Four sample mass-radius relations are also shown: cold-hydrogen (blue dashed line), Earth-like rocky (green dash-dotted line), pure-iron (black dotted line), and pure rocky (solid crimson line) planets \citep{2010ApJ...712L..73M,2014ApJS..215...21B}.}
    \label{fig1}
\end{figure}

\section{Introduction}
Our comprehension of new worlds beyond the Solar System, known as exoplanets, their population, and diversity come largely from the latest generation of modern satellites. The Kepler space mission, the Transiting Exoplanet Survey Satellite, the James Webb Space Telescope, and many ground-based observatories make important contributions to detecting and characterizing exoplanets \citep{2007PASP..119..923P,2010Sci...327..977B,2014PASP..126.1134B}. The data generated by these state-of-the-art instruments are now available to everyone. Researchers skilled in data science, data analytics, or machine learning (ML) and neural network techniques study and analyze these data to predict, identify, characterize, and classify the exoplanets \citep{2019A&A...626A..21A,2019MNRAS.487.5062M,2020EPSC...14..833B,2020AJ....159...41T,2021MNRAS.504.5327A,2021A&A...649A..26L,2021A&A...655A..66L,2021PASA...38...15M,2021A&A...656A..73S,2021MNRAS.507.2154V,2023A&A...670A..68M,2023NatAs...7....8M}. In addition, the observational data are not only used to study exoplanets but they are also applied to peruse entire planetary science. As many planets are found around other stars, they have provided us with an opportunity to understand the main ways of planet formation and evolution and to put our Solar System in a broader context \citep{2018MNRAS.473..784K,2020apfs.book.....A,2020AJ....159..281G,mishra2023framework}.

At this paper's writing, more than 5,000 exoplanets have been discovered, and thousands of candidated are yet to be confirmed. Transit and radial velocity are two fundamental methods to discover exoplanets and determine their main parameters. The transit method regularly observes the small fraction of the star's light blocked by a transiting planet. Observing this light decrement makes it possible to calculate the planet's radius. On the other hand, using the radial velocity method, the slight movement of a star caused by an orbiting planet is measured, and the planet's mass is obtained. However, not all discovered planets have measured mass and radius as two important physical properties \citep{2010exop.book.....S,2018haex.bookE.117D}.

There is a clear correlation between the radius and mass of a planet that can be described with a polytropic relation \citep{1993RvMP...65..301B,2000ARA&A..38..337C}. Many works have investigated the relationship between planetary parameters, particularly mass, and radius, and to deduce the composition and structure of exoplanets \citep{2007ApJ...669.1279S,2012ApJ...744...59S,2020A&A...634A..43O}. \citet{2013ApJ...768...14W} divided the planets into two groups, those with masses greater than and lower than 150$M_{\oplus}$, and presented a power law relation for the mass-radius distribution of each group. \citet{2017AA...604A..83B} revised this mass breakpoint to $124\pm7M_{\oplus}$ and proposed $R_{p}\propto M^{0.55\pm0.02}$ for small planets and $R_{p}\propto M^{0.01\pm0.02}$ for large planets. Assuming a power law description of the mass-radius relation, for the first time, a probabilistic model for planets with radii lower than 8$R_{\oplus}$ was presented by \citet{2016ApJ...825...19W}. \citet{2017ApJ...834...17C} implemented this idea to an extended dataset, forecasting the mass or radius of planets. Moreover, they calculated the forecasted mass for $\sim$7000 Kepler Objects of Interest \citep{2018MNRAS.473.2753C}.

Most previous works use a power law model to explore the mass-radius relation and have assumptions that are not pliable enough to consider principal attributes in such diagrams. Consequently, \cite{2018ApJ...869....5N} developed a non-parametric approach using a sequence of Bernstein polynomials and the sample of \citet{2016ApJ...825...19W}. The same method was used in follow-up work to analyze the mass-radius relation of exoplanets orbiting M dwarfs \citep{2019ApJ...882...38K}.

The correlation between physical parameters in planetary systems is not limited to the planet's mass and radius. It has been demonstrated that the radius of a giant planet is related to other parameters such as the orbital semi-major axis, the planetary equilibrium temperature, the tidal heating rate, and the stellar irradiation and metallicity \citep{2006A&A...453L..21G,2007ApJ...659.1661F,2012A&A...540A..99E}. \citet{2002ApJ...568L.113Z} reported a possible correlation between the mass and period of an exoplanet. They also showed that planets revolving around a binary host star might have an opposite correlation. \citet{2014ApJ...783L...6W} studied a restricted dataset containing 65 exoplanets smaller than 4$R_{\oplus}$ with orbital periods shorter than 100 days. They showed that planets smaller than 1.5$R_{\oplus}$ are consistent with a positive linear density–radius relation, but for planets larger than 1.5$R_{\oplus}$, density decreases with radius. \citet{2015ApJ...810L..25H} presented the mass-density relationship in a logarithmic space for objects ranging from planets ($M\approx0.01M_{J}$) to stars ($M>0.08M_{\odot}$). They divided the mass-density distribution into three regions based on changes in the slope of the relationship and introduced a new definition for giant planets.

\citet{2016arXiv160700322B} used a Random Forest regression model to evaluate the influence of different physical parameters on planet radii. Applying this model to different groups of giant planets, they found that the planet's mass and equilibrium temperature has the greatest effect on determining the radius of a hot-Saturn ($0.1<M_{p}<0.5M_{J}$). They also showed that the equilibrium temperature is more important for more massive planets. Moreover, \citet{2019A&A...630A.135U} (hereafter, \citetalias{2019A&A...630A.135U}) introduced Random Forest as a promising algorithm for obtaining exoplanet properties. They used Random Forest to predict the exoplanet radii based on several planetary and stellar parameters. Similar to previous results, an exoplanet's mass and equilibrium temperature were the fundamental parameters.

As the number of discovered exoplanets rapidly increases, ML techniques can be used to investigate correlations between planets and their host stars. In this study, we implement various ML algorithms to find the potential relationships between physical parameters in exoplanet systems. The Markov Chain Monte Carlo (MCMC) \citep[see][]{goodman10,emcee3} is used to quantify the uncertainties of the best-fit parameters. In addition, different ML clustering algorithms are used to group the exoplanets and study their properties. We organize this paper as follows: In Sect.~\ref{data}, the sample data and methods of pre-processing, clustering, and modeling are introduced. Section~\ref{result} presents the results. In Sect.~\ref{conclusion}, we summarise the main results and conclusions.

\section{Data and methods}\label{data}
\subsection{Dataset}
We use the NASA Exoplanet Archive\footnote{\url{https://exoplanetarchive.ipac.caltech.edu/}} and the Extrasolar Planets Encyclopedia\footnote{\url{http://exoplanet.eu/}} to extract the data of exoplanets\footnote{The data was last extracted on March 22, 2022.}. These two catalogs are comprehensive, up-to-date, and available to the public, and also provide access to relevant publications \citep{2011epsc.conf....3S,2013PASP..125..989A}. There are 762 confirmed exoplanets with reported physical parameters, including the orbital period ($P$) and eccentricity ($e$), planetary mass ($M_{p}$) and radius ($R_{p}$), and the stellar mass ($M_{s}$), radius ($R_{s}$), metallicity (Fe/H), and effective temperature ($T_{\text{eff}}$). Note that exoplanets with only a minimum mass are not considered. Among these 762 exoplanets, six have been discovered by the radial velocity method, and each of the imaging and transit timing variation methods has identified only one exoplanet. In general, most exoplanets have been discovered by observing the slight decrease in brightness of the host star caused by the transit of a planet in front of it. Using NASA's Planetary Fact Sheet\footnote{\url{https://nssdc.gsfc.nasa.gov/planetary/factsheet/}}, we add the eight Solar System planets to the sample. Overall, the dataset contains 770 planets. Fig.~\ref{fig1} shows the radius of planets plotted as a function of mass and color coded by orbital period. It is separated into four groups based on detection methods: transit, radial velocity, transit timing variation, and imaging. Distributions of cold-hydrogen, Earth-like rocky (32.5\% Fe+67.5\% MgSiO3), pure-iron (100\% Fe), and pure-rock (100\% MgSiO3) planets are also illustrated \citep{2010ApJ...712L..73M,2014ApJS..215...21B}. We should note that, like other observational datasets, our sample is also affected by detection biases. A dataset containing planets whose radius and mass have been measured suffers from the detection limits of both radial velocity and transit methods. Thus, it is impossible to draw a reliable conclusion about the occurrence of planets using our dataset.

\subsection{Data pre-processing}\label{pre}
We aim to predict the planetary radius as the target variable, using other physical parameters as features. As the parameters have different ranges, we transfer them to a logarithmic space. In data analysis, the ultimate results might be affected by some unreliable measurements in the sample. In machine learning and statistics, there are diverse methods to detect these unusual observations in a dataset. We choose the Local Outlier Factor (LOF) method to identify and remove observations with abnormal distances from other values. The technique is often used with multidimensional datasets, like our eight-dimensional one, which has different densities and types of outliers. The LOF uses two hyperparameters: neighborhood size ($k$), which defines the neighborhood for local density calculation, and contamination ($c$), which specifies the proportion of outliers in the dataset \citep{breunig2000lof,chandola2009anomaly}. By choosing $k=20$ and $c=0.05$, the LOF method is run twice: once for all parameters including $P$, $e$, $M_{p}$, $R_{p}$, $M_{s}$, $R_{s}$, Fe/H, and $T_{\text{eff}}$, then for planetary mass and radius, which are known to be highly correlated. Altogether, the LOF detects 76 data points as outliers. Appendix~\ref{appA} describes the process of identifying outliers in detail.

Choosing appropriate features plays a vital role in building an efficient ML model. Adding extra variables or those highly correlated with each other may reduce the overall predictive ability of the model and lead to wrong results. Feature selection (hereafter FS) methods rank features based on their usefulness and effectiveness in making predictions. The FS methods can be divided into three groups: filter, wrapper, and embedded methods \citep{guyon2008feature,chandrashekar2014survey,jovic2015review,brownlee2016machine}. In filter methods, features are filtered independently of any induction algorithm and based on some performance evaluation metrics calculated directly from the data \citep{sanchez2007filter,cherrington2019feature}.
In contrast, the selection process in wrapper methods is based on the performance of a specific ML algorithm operating with a subset of features \citep{ferri1994comparative,kohavi1997wrappers,hall1999feature}. Embedded methods combine the qualities of both filter and wrapper methods. They perform the FS in the training process and are usually specific to given learning machines \citep{lal2006embedded,bolon2013review}. We use five FS methods to identify the most important features in our dataset: Spearman's rank correlation test as a filter method, the Backward Elimination and Forward Selection as two wrapper methods, and the CART (classification and regression trees) and XGBoost (extreme gradient boosting) as two Embedded methods.

\subsection{Clustering}
Clustering as an unsupervised ML task involves grouping each data point with a specific type. In theory, data points belonging to a particular group should have similar properties \citep{xu2005survey,kaufman2009finding,2012arXiv1205.1117S}. Data clustering algorithms can be divided into hierarchical and partitional groups. Hierarchical algorithms find clusters using previously established clusters, while partitional algorithms find all clusters at once \citep{kononenko2007chapter,2012arXiv1205.1117S}. We aim to group planets into distinct, non-overlapping clusters, similar to exclusive clustering \citep{1988acd..book.....J}. We use ten ML clustering algorithms to include a wide range of clustering methods and examine their performance in exoplanet data. Since many diverse exoplanets have been discovered, implementing clustering algorithms can find potential exoplanet groups to investigate their characteristics. The algorithms are available in the Scikit-learn software ML library \citep{scikit-learn} and are as follows: Affinity Propagation, BIRCH (balanced iterative reducing and clustering using hierarchies), DBSCAN (density-based spatial clustering of applications with noise), Gaussian Mixture Model, Hierarchical Clustering, K-Means, Mean Shift, Mini-Batch K-Means, OPTICS (ordering points to identify the clustering structure), and Spectral Clustering \citep{davies1979cluster,ester1996density,zhang1996birch,ankerst1999optics,halkidi2001clustering,comaniciu2002mean,2007Sci...315..972F,von2007tutorial,schubert2017dbscan}.

BIRCH, Gaussian Mixture Model, Hierarchical Clustering, K-Means, Mini-Batch K-Means, and Spectral Clustering are algorithms that do not learn the number of clusters ($K$) from data. Therefore, we first perform the Elbow and Silhouette methods to find the optimal number of clusters. The Elbow method runs the K-Means clustering algorithm for $K$ values. Then, for each $K$, it computes the sum of squared distances (SSD) between data points and their assigned cluster centroids and uses them to propose an optimal number of clusters. The Silhouette method determines the degree of separation between clusters by choosing a range of $K$ values and calculating a coefficient for each $K$. The silhouette coefﬁcient for a particular data point is calculated by $(b^{i} - a^{i})/{max(a^i, b^i)}$. Here, $a^{i}$ represents the average distance from all data points in the same cluster, whereas $b^{i}$ is the average distance from data points that belong to the closest cluster. Provided that the sample is on or near the decision boundary between two neighboring clusters, the silhouette coefficient becomes 0. A coefficient close to +1 indicates that the sample is far from neighboring clusters. A negative coefficient value indicates that samples may have been assigned to the wrong cluster \citep{rousseeuw1987silhouettes,2012arXiv1205.1117S}.

\begin{figure*}
	\includegraphics[width=0.85\paperwidth]{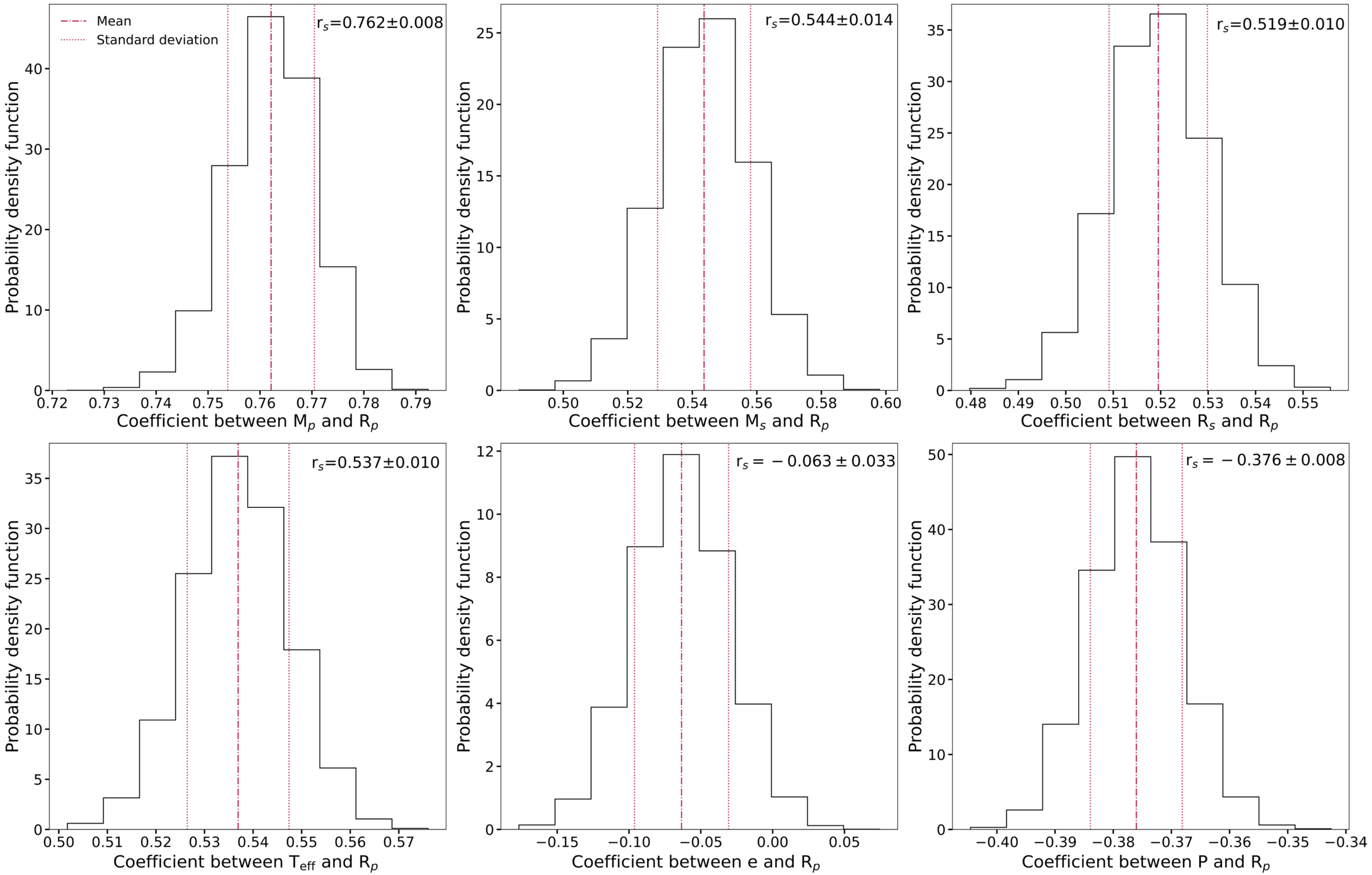}
    \caption{Distribution of correlation coefficient ($r_{s}$) between planetary radius ($R_{p}$) and other physical parameters obtained by the Monte Carlo analysis. Parameters are planetary mass ($M_{p}$), stellar mass ($M_{s}$), radius ($R_{s}$), and effective temperature ($T_{\text{eff}}$), and orbital eccentricity ($e$) and period ($P$). Stellar metallicity (Fe/H), which displays a p-value greater than 0.1, has been excluded. The red dash-dotted line is the mean, and two red dotted lines represent uncertainties around it. The mean and uncertainty values are presented in the upper right corner of each panel. The absolute value of the coefficients determines the strength of the relationship. The larger the number, the stronger the relationship. It marks $M_{p}$ as the most and $e$ as the least relevant parameters to the planetary radius.}
    \label{fig2}
\end{figure*}

\subsection{Modeling}
In machine learning, different algorithms allow machines to learn information from a given dataset, uncover relationships, and make predictions \citep{brownlee2016machine,brownlee2016machine2}. In our case, we apply ML models to predict planet radii when other parameters are given. It is also possible to see how efficiently each parameter uses these models. The algorithms used to perform regression tasks are as follows: Decision Tree, K-Nearest Neighbors, Linear Regression, Multilayer Perceptron, M5P, and Support Vector Regression (SVR) \citep{hinton1990connectionist,quinlan1992learning,quinlan1993program,hastie2009elements,chang2011libsvm}. Bootstrap Aggregation and Random Forest are also used as ensemble algorithms that combine the predictions from multiple models \citep{breiman1996bagging,breiman2001random}. These algorithms are available in the Weka tool environment \citep{witten2005practical,hall2009weka}.

Linear Regression and M5P are two algorithms that can extract parametric equations. Linear Regression algorithm fits a linear model to the entire data. M5P performs a multiple linear regression model. This tree-based algorithm allocates linear regressions at the terminal nodes. It divides the entire dataset into several smaller subsets and fits a linear model to each subset \citep{quinlan1992learning}. They both, consequently, result in a basic linear equation like Eq.~\ref{eq1}, where $A$ and $C$ are fitting parameters, $Y$ is the dependent variable, $X$ is the independent variable, and $N$ represents the total number of independent variables. In our case, $Y$ is the planet’s radius, and $X$ represents other physical parameters. For these two algorithms, we use the MCMC to quantify the uncertainties of the best-fit parameters \citep{goodman10,emcee3}.

\begin{equation}
Y=C+\sum_{i=1}^{N}A_{i}X_{i}.
\label{eq1}
\end{equation}

To evaluate the quality of the predicted radius ($R_{pre}$) compared to the observed radius ($R_{obs}$) and to compare the efficiency of the models, root means square error (RMSE), mean absolute error (MAE), and coefficient of determination ($\rho^{2}$) are calculated. Eq.~\ref{eq2} to~\ref{eq4} define RMSE, MAE, and $\rho^{2}$, respectively, where $R_{mean}$ is the mean of the $R_{obs}$ values and $n$ represents the total number of samples. Lower values of RMSE and MAE and higher values of $\rho^{2}$ indicate better accuracy of the models. It should be noted that hyperparameters specific to each model are tuned to have the best performance. Also, a 10-fold cross-validation procedure, as a data resampling method, is used to evaluate the performance of models. Furthermore, each algorithm is executed for original and logarithmic datasets to understand the effect of data re-scaling.

\begin{equation}
RMSE=\sqrt{\sum_{i=1}^{n}\frac{(R_{obs}-R_{pre})^2}{n}}.
\label{eq2}
\end{equation}

\begin{equation}
MAE=\frac{1}{n}\sum_{i=1}^{n}\mid R_{obs}-R_{pre}\mid.
\label{eq3}
\end{equation}

\begin{equation}
\rho^{2}=1-\frac{\sum_{i=1}^{n}(R_{obs}-R_{pre})^2}{\sum_{i=1}^{n}(R_{obs}-R_{mean})^2}.
\label{eq4}
\end{equation}

\section{Results}\label{result}
Before applying the clustering and predictive methods, we use the LOF algorithm to identify outliers in the dataset containing 770 data points. The LOF algorithm marks 76 data points as outliers, resulting in a study dataset of 694 planets. Appendix~\ref{appA} describes finding outliers and their effect on prediction accuracy. In general, we find that all regression models perform poorly considering the outliers in the dataset. Additionally, the effect of data re-scaling on planetary radius predictions is discussed in Appendix~\ref{appB}. As a result, regression algorithms provide better results on a logarithmic scale.

\subsection{Feature importance}
It has been known that having inefficient and unnecessary features can cause declination in the performance of an ML model and lead to inaccurate results \citep{chandrashekar2014survey}. We apply five FS methods to find and eliminate the least important planetary, stellar, and orbital parameters in predicting planet radii. They are the Spearman’s rank correlation, Backward Elimination, Forward Selection, CART, and XGBoost. Features are orbital period ($P$), eccentricity ($e$), planetary mass ($M_{p}$), stellar mass ($M_{s}$), stellar radius ($R_{s}$), metallicity (Fe/H), and effective temperature ($T_{\text{eff}}$), and the target variable is planetary radius ($R_{p}$).

\begin{figure}
	\includegraphics[width=\columnwidth]{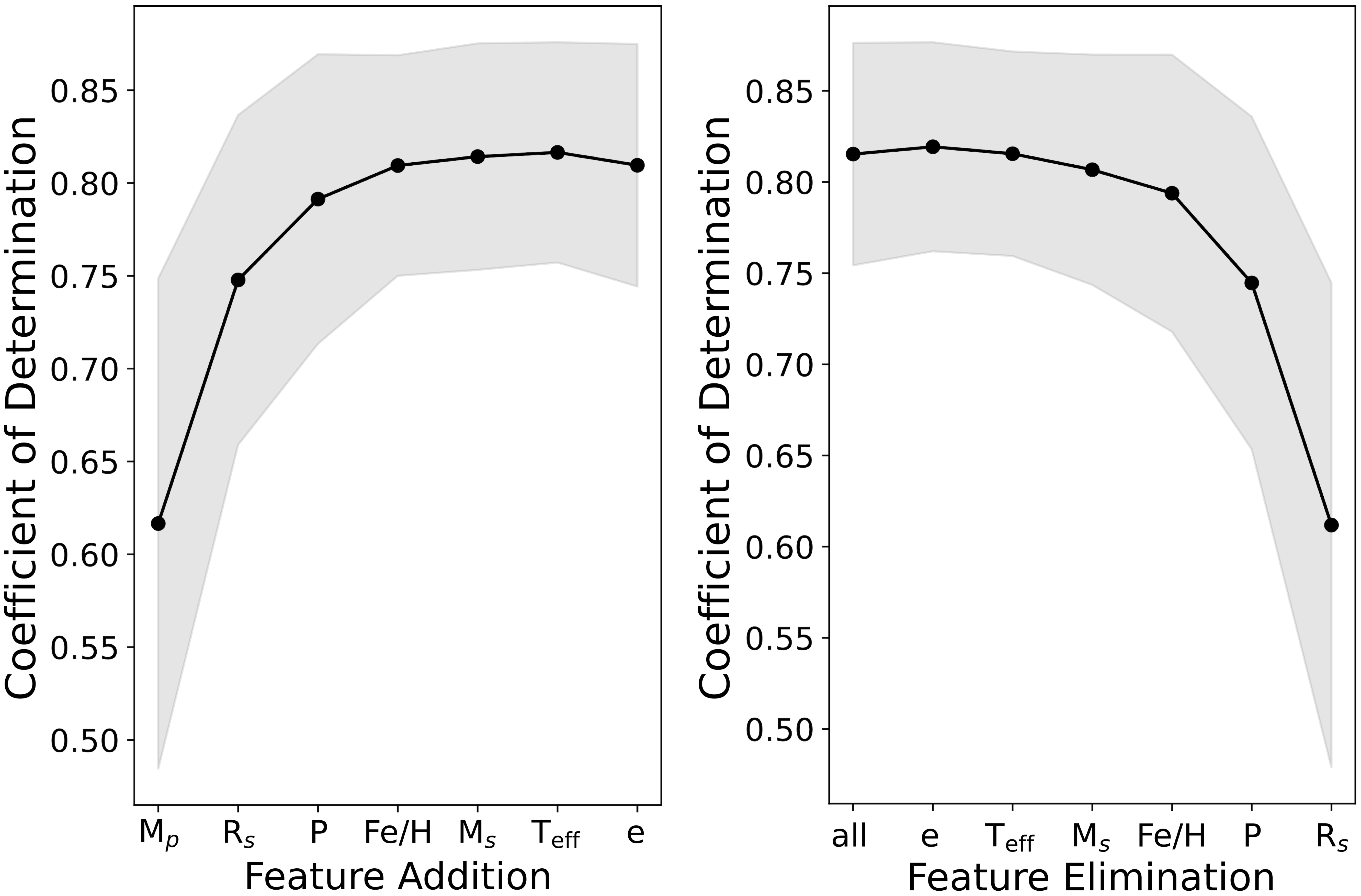}
    \caption{Wrapper FS methods. Left: coefficients of determination ($\rho^{2}$) against feature sets for the Forward Selection technique. In the first step, it determines $M_{p}$ as the best feature, and in each following iteration, the best remaining feature is added to the set. Right: $\rho^{2}$ values against feature sets for the Backward Elimination technique. Unlike Forward Selection, it starts with all features and removes the worst one at each step. Features are planetary mass ($M_{p}$) and radius ($R_{p}$), orbital period ($P$) and eccentricity ($e$), and the stellar mass ($M_{s}$), radius ($R_{s}$), metallicity (Fe/H), and effective temperature ($T_{\text{eff}}$). Gray areas demonstrate standard errors. Both methods highlight $M_{p}$, $P$, and $R_{s}$ as three important parameters in predicting the planetary radius.}
    \label{fig3}
\end{figure}

Spearman’s rank correlation ($r_{s}$) is a number between –1 and 1 that measures the monotonic correlation between two variables. This filter method reveals parameters that are strongly correlated with planetary radius: $M_{p}$ with a coefficient of 0.780 has the highest correlation, followed by $M_{s}$ with $r_{s}=0.590$. Furthermore, $T_{\text{eff}}$ with a coefficient of 0.568 and $R_{s}$ with a coefficient of 0.552 are in third and fourth place, respectively. Orbital period with $r_{s}=-0.389$ and eccentricity with $r_{s}=-0.151$ are two features that have a negative correlation with $R_{p}$. In addition, highly correlated stellar parameters ($R_{s}$, $M_{s}$, and $T_{\text{eff}}$) are also indicated by coefficients greater than 0.820. We should note that the estimated p-values are less than 0.001, which indicates strong certainty in the results. As an exception, the p-value corresponding to the coefficient between Fe/H and $R_{p}$ ($r_{s}=-0.037$) is greater than 0.1, which is statistically uncertain. To calculate uncertainties in the value of correlation coefficients, we apply the Monte Carlo error analysis, considering errors in each measurement \citep{2015ascl.soft04008C}. Fig.~\ref{fig2} illustrates the distribution of $r_{s}$ for each feature except Fe/H, which has a p-value greater than 0.1. By taking the standard deviation of distributions, we report the mean and uncertainty values in the upper right corner of each panel. As it is seen, the distributions of Spearman correlation coefficients for $M_{p}$, $M_{s}$, $R_{s}$, and $T_{\text{eff}}$ are constrained to positive values. In contrast, the distribution for $P$ is limited to negative values. In the case of $e$, even though the mean of the distribution is slightly negative, it is exceptionally wide, taking both negative and positive values. This more extensive distribution results from higher amounts of error in eccentricity measurement.

Forward Selection and Backward Elimination are two wrapper FS methods. The procedure for Forward Selection starts with an empty set of features. Then, the best feature is determined and added to the set by applying a Random Forest regressor. In each subsequent iteration, the best remaining feature is determined and added until a complete set of features is reached. In contrast, Backward Elimination starts with a complete set of features and, at each step, eliminates the worst feature remaining in the set. Using a 10-fold cross-validation method, $\rho^{2}$ values, and corresponding standard errors are calculated for each step and shown in Fig.~\ref{fig3}, where the left-hand panel is Forward Selection, and the right-hand panel is Backward Elimination. Both methods highlight $M_{p}$, $P$, and $R_{s}$ as three important parameters.

\begin{figure}
	\includegraphics[width=\columnwidth]{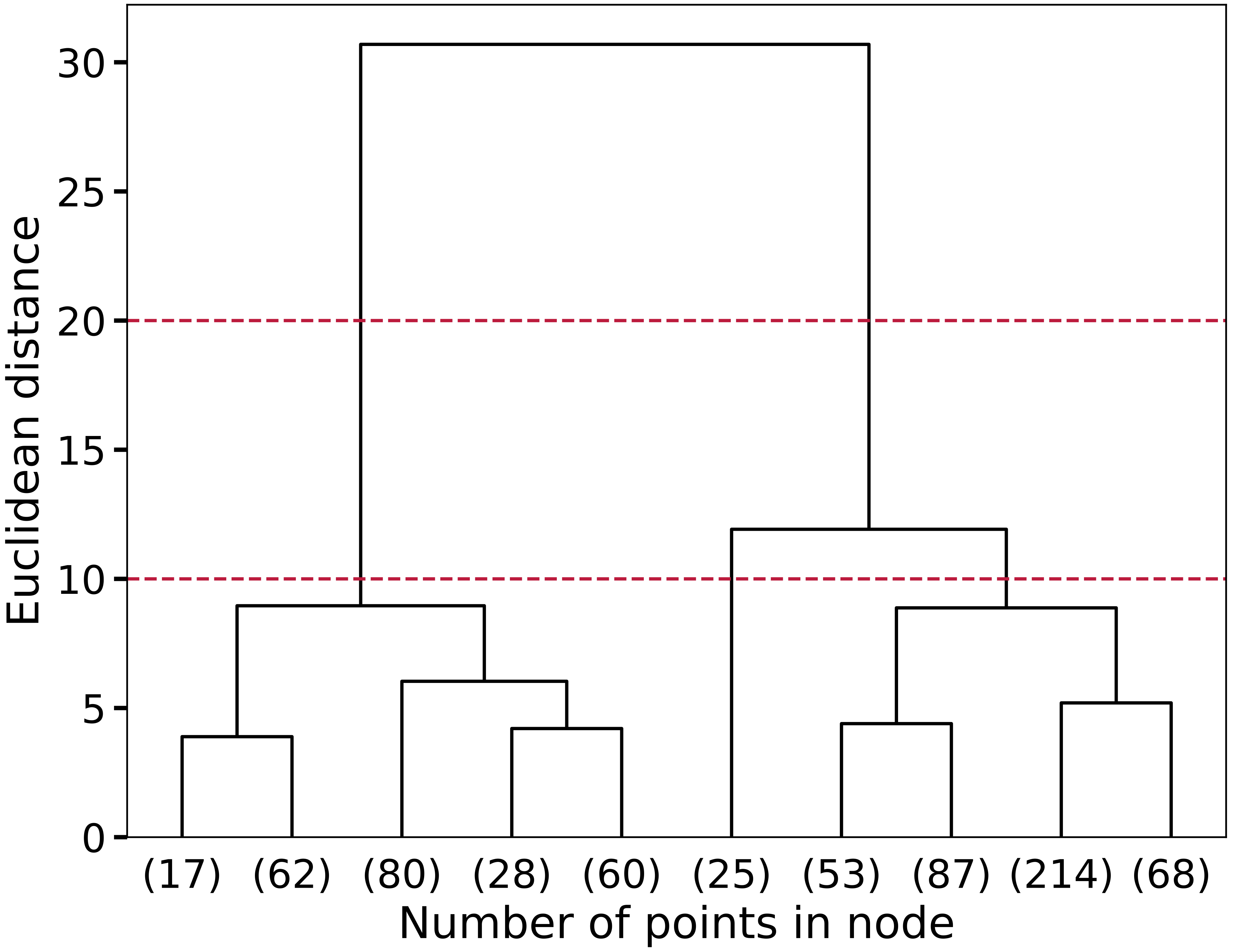}
    \caption{Hierarchical Clustering dendrogram. The greater the height of the vertical lines, the greater the distance between the clusters. Two red dashed lines are distance thresholds. The number of vertical lines intersecting the threshold line indicates the number of clusters. The larger threshold results in two clusters, while the smaller threshold results in three clusters. Due to illustration purposes, lower sequences have not been shown.}
    \label{fig4}
\end{figure}

\begin{figure}
	\includegraphics[width=\columnwidth]{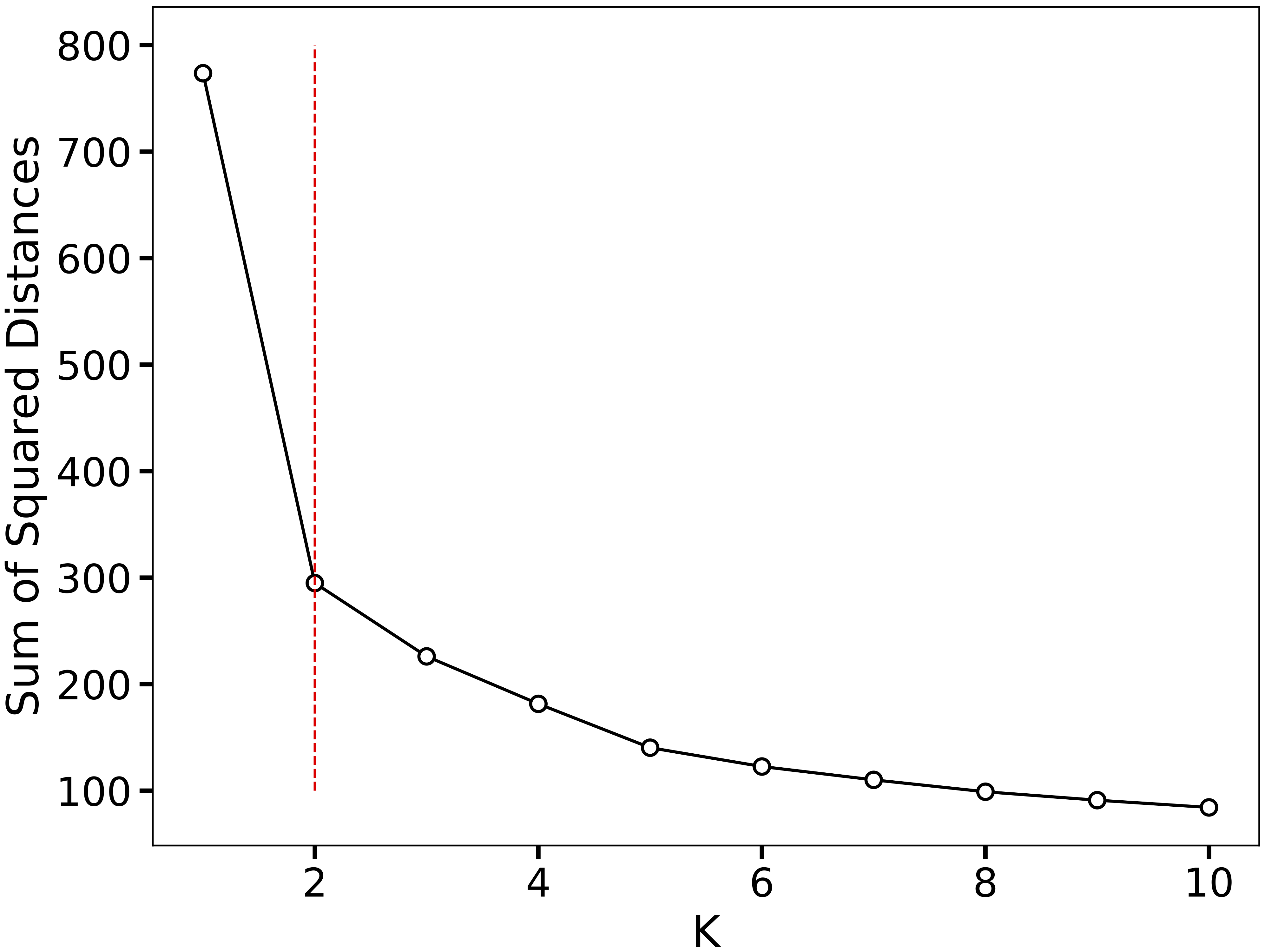}
    \caption{Sum of squared distances (SSD) between data points and their assigned cluster centroids against the number of clusters ($K$), calculated by the Elbow method. The red vertical dashed line corresponds to $K$=2, where the curve starts to flatten out. This point is chosen as the optimal number of clusters.}
    \label{fig5}
\end{figure}

\begin{figure}
	\includegraphics[width=\columnwidth]{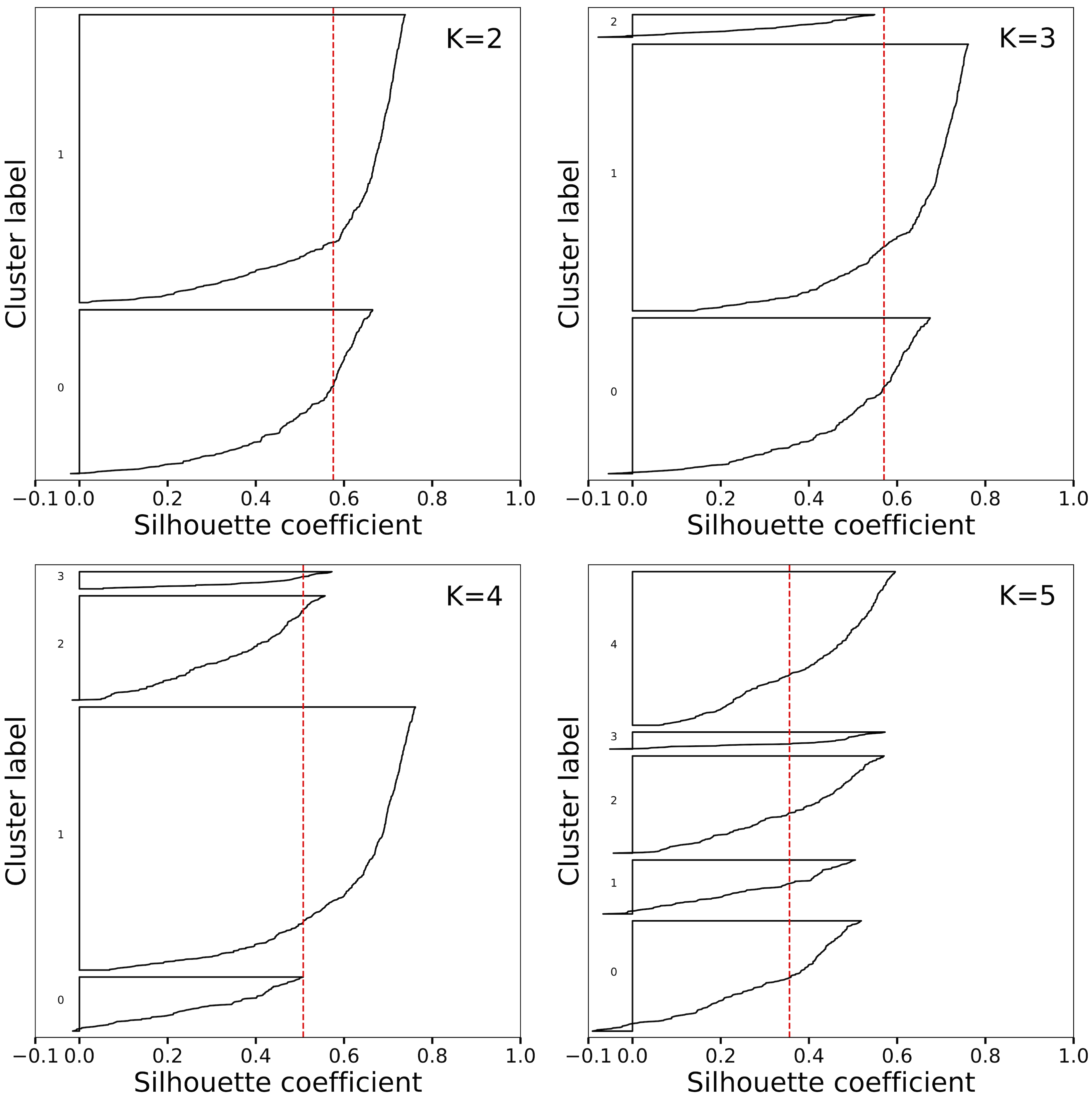}
    \caption{Silhouette plots for different numbers of clusters ($K$). The thickness of the plots indicates the cluster size and red vertical dashed lines represent the corresponding average Silhouette coefficients. If the sample is far from neighboring clusters, the coefficient becomes close to +1. The coefficient can be negative if the sample is assigned to the wrong cluster. Providing that the sample is on or near the decision boundary between two neighboring clusters, the coefficient becomes 0. Due to clusters with lower-than-average Silhouette scores and wide fluctuations in the size of plots, $K$=3, 4, and 5 are not appropriate. Scatter plots for each Silhouette plot are shown in Fig.\ref{fig7}.}
    \label{fig6}
\end{figure}

Applied embedded methods include CART, which uses a decision tree regressor, and XGBoost, which implements a gradient boosting trees algorithm. These techniques score features based on their importance in computing the target variable. The ranking (and scores) obtained by the CART method are as follows: $M_{p}$ (0.905), $P$ (0.031), $R_{s}$ (0.026), $M_{s}$ (0.012), $T_{\text{eff}}$ (0.011), $e$ (0.008), and Fe/H (0.006). The XGBoost method ranks (and scores) features as follows: $M_{p}$ (0.851), $R_{s}$ (0.038), $P$ (0.029), $M_{s}$ (0.025), $T_{\text{eff}}$ (0.024), Fe/H (0.021), and $e$ (0.012). Similar importance is assigned to all features except planetary mass by CART and XGBoost. Finally, we conclude that a set of features including $M_{p}$, $P$, and one of the stellar parameters ($M_{s}$, $R_{s}$, or $T_{\text{eff}}$) works well. Therefore, we select planetary mass, orbital period, and stellar mass as the main features.

\subsection{Clusters}
FS methods highlight planetary mass, stellar mass, and orbital period as vital features to estimate planet radii. We use ML clustering algorithms to investigate potential groups of exoplanets in a four-dimensional logarithmic space consisting of planetary mass and radius, stellar mass, and orbital period. The algorithms are Affinity Propagation, BIRCH, DBSCAN, Gaussian Mixture Model, Hierarchical Clustering, K-Means, Mean Shift, Mini-Batch K-Means, OPTICS, and Spectral Clustering.

\subsubsection{Number of clusters}\label{number}
When the number of clusters ($K$) is not known in advance, as in our case, hierarchical clustering is an appropriate technique to adopt \citep{landau2011cluster}. Using the Hierarchical Clustering algorithm as a distance-based method, one can have an assumption of $K$. Its agglomerative algorithm assigns each data point to an individual partition; then, at each iteration, the closest pair of partitions are merged until all data belong to a single partition. Fig.~\ref{fig4} shows the Hierarchical Clustering dendrogram, which records the sequences of merges. The greater the height of the vertical lines in the dendrogram, the greater the distance between the clusters. Clusters can be defined by trimming a dendrogram with a distance threshold. However, there is no universal method for setting thresholds. In general, a distance threshold on the dendrogram is set to intersect the longest vertical line. The number of vertical lines intersecting the threshold line indicates the $K$ value. Two distance thresholds are set to the dendrogram of our dataset (red dashed lines in Fig~\ref{fig4}). The larger threshold results in two clusters with a Euclidean distance of 30.6, while the smaller threshold splits the larger cluster into two parts with a distance of 11.9 between them.

For algorithms that do not learn the $K$ from data, we use Elbow and Silhouette methods to find the optimal value of $K$ and use it as an input parameter in clustering algorithms. The Elbow method performs the K-Means clustering algorithm for different values of $K$. Then, it calculates the sum of squared distances (SSD) between data points and their cluster centroids each time. Fig.~\ref{fig5} demonstrates SSD values as a function of $K$. As shown, the curve starts to flatten out and form an elbow shape in $K$=2, chosen as the optimal number of clusters.

\begin{figure*}
	\includegraphics[width=0.82\paperwidth]{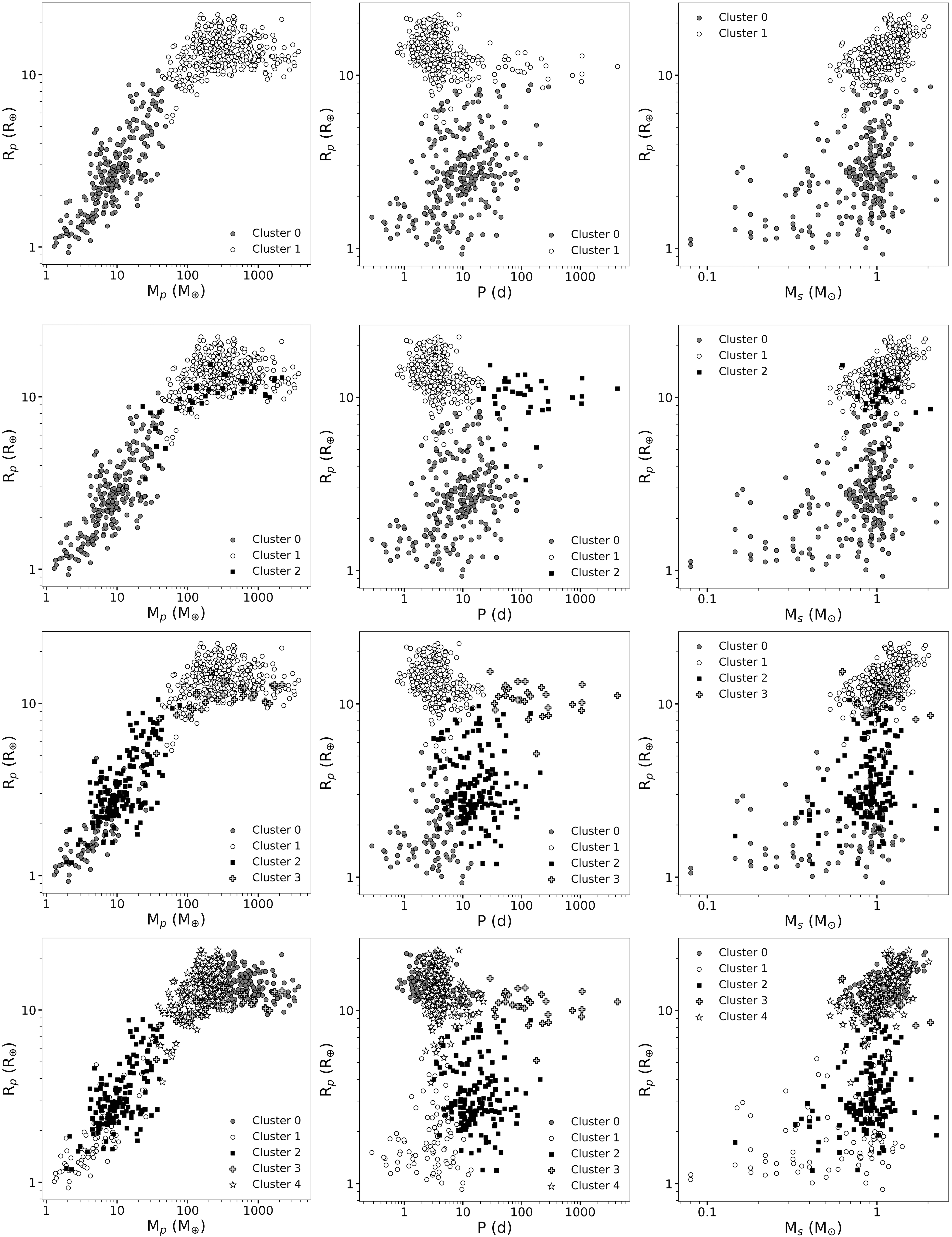}
    \caption{Matrix of scatter plots corresponding to Fig.~\ref{fig6}. Rows 1, 2, 3, and 4 represent the number of clusters ($K$) equal to two, three, four, and five, respectively. Columns 1, 2, and 3 illustrate the distribution of planetary radius ($R_{p}$) versus the planet’s mass ($M_{p}$), orbital period ($P$), and star’s mass ($M_{s}$), respectively. Gray-filled circles, white circles, black-filled squares, light gray-filled pluses, and white stars represent members of clusters 0, 1, 2, 3, and 4, respectively.}
    \label{fig7}
\end{figure*}

Furthermore, the average Silhouette width can evaluate clustering reliability and may be used to estimate $K$ value \citep{rousseeuw1987silhouettes}. The Silhouette method computes a coefficient for different values of $K$ and uses it to determine the degree of separation between clusters. The coefficient becomes negative if the sample is assigned to the wrong cluster. Provided that the sample is far from neighboring clusters, the coefficient will be close to +1. If the sample is on or near the decision boundary between two neighboring clusters, the coefficient becomes 0. Fig.~\ref{fig6} presents the Silhouette plots for $K$=2, 3, 4, and 5. In this figure, the thickness of the plots represents the cluster size, and each red vertical dashed line belongs to an average Silhouette score. $K$=3, 4, and 5 are not appropriate due to clusters with lower-than-average Silhouette scores and wide fluctuations in the size of plots. Like the Elbow method, this method proposes $K$=2 as the proper number of clusters. Fig.~\ref{fig7} illustrates a matrix of scatter plots for Silhouette plots, where each row has been assigned to a specific $K$, and columns 1, 2, and 3 show the distribution of planetary radius versus planet’s mass, orbital period, and star’s mass, respectively. Like exclusive clustering algorithms where each data point belongs exclusively to one cluster \citep{1988acd..book.....J}, we aim to group planets into distinct non-overlapping clusters. On the one hand, if $K=2$ (see the first row in Fig.~\ref{fig7}), two clusters are almost well-separated in all three spaces, with only a few data points overlapping. On the other hand, when planets are divided into more than two clusters (see the second, third, and fourth rows in Fig.~\ref{fig7}), the members of the clusters become less distinguishable from each other.

\begin{figure}
\includegraphics[width=\columnwidth]{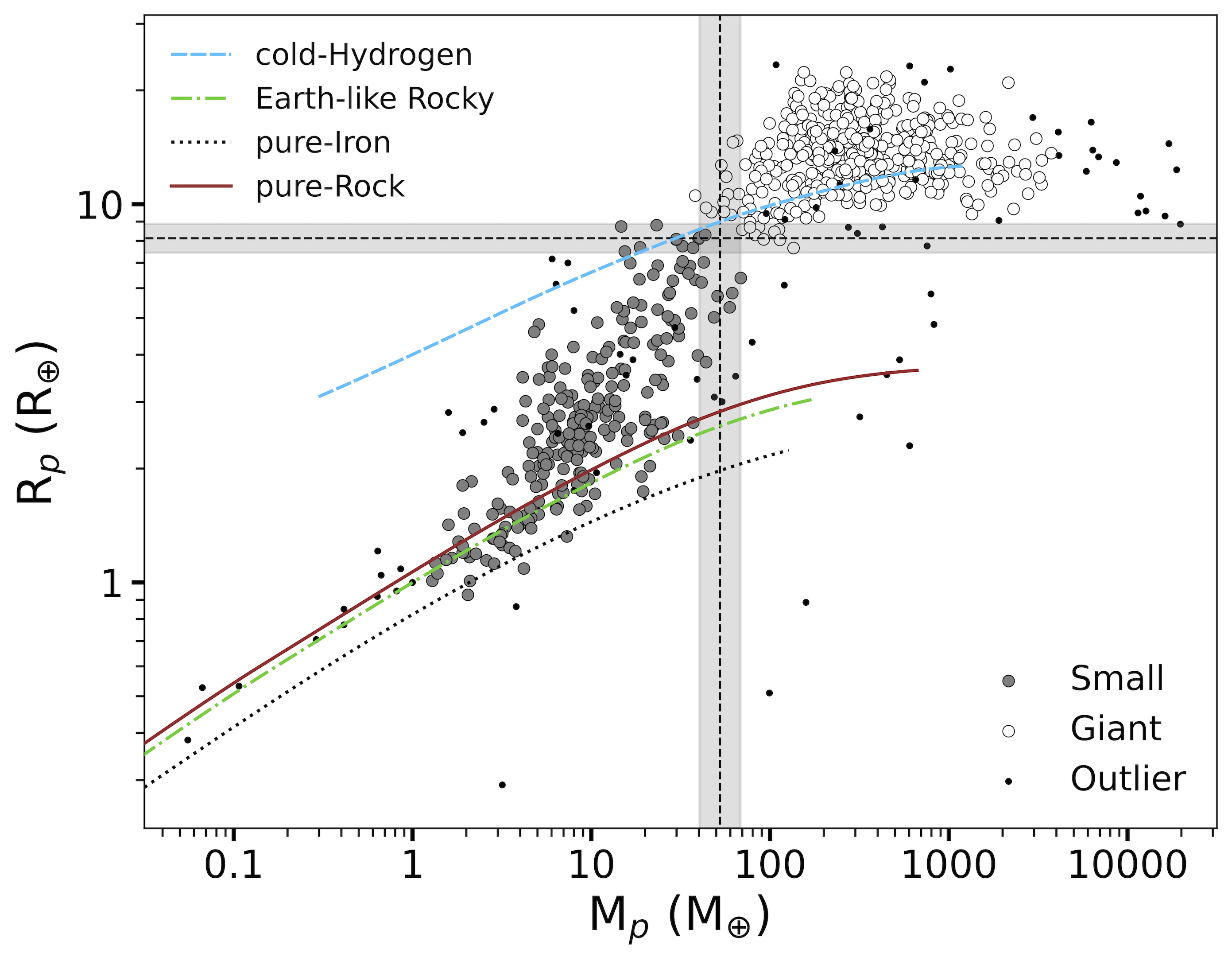}
\caption{The mass-radius distribution of clustered data. Data are separated into two classes using $R_{p}=8.13R_{\oplus}$ (horizontal dashed line) and $M_{p}=52.48M_{\oplus}$ (vertical dashed line). Gray areas demonstrate mean errors of mass and radius. Exoplanets with $R_{p}\leq8.13R_{\oplus}$ and $M_{p}\leq52.48M_{\oplus}$ are defined as small planets (gray circles), and those with $R_{p}>8.13R_{\oplus}$ and $M_{p}>52.48M_{\oplus}$ as giant planets (white circles). There are 254 small planets and 440 giant planets. Black dots are outlier planets found by the LOF method (see Appendix~\ref{appA}). Four iso-density curves are also drawn: cold-hydrogen (blue dashed line), Earth-like rocky (green dash-dotted line), pure-iron (black dotted line), and pure rocky (solid crimson line) planets \citep{2010ApJ...712L..73M,2014ApJS..215...21B}.}
\label{fig8}
\end{figure}

For $K=3$, planets with a longer orbital period are defined as a new cluster (black-filled squares). Although three clusters are separated in the $R_{p}$-$P$ space, in the $R_{p}$-$M_{p}$ and $R_{p}$-$M_{s}$ spaces, most members of cluster 2 are distributed over the other two clusters, especially cluster 1. Likewise, in the case of $K=4$, the cluster separation is better in the $R_{p}$-$P$ space than in the other two spaces, where clusters overlap.

For $K=5$, although clusters 0 and 4 are well distinguished in the planet’s mass-radius distribution, they overlap a lot in the $R_{p}$-$P$ and $R_{p}$-$M_{s}$ spaces. In the $R_{p}$-$P$ space, members of clusters 0, 1, 2, and 3 are almost separated; nevertheless, members of cluster 4 extremely overlap with those of cluster 0. In addition, cluster 3 members overlap with members of clusters 0 and 4 in the $R_{p}$-$M_{p}$ and $R_{p}$-$M_{s}$ spaces.

Consequently, we choose $K=2$ based on the results from the Hierarchical, Elbow, and Silhouette methods and the idea that planetary clusters are separate groups that do not overlap. Similarly, the Affinity Propagation and Mean Shift algorithms, which do not need to specify the number of clusters, give two clusters. Another point that should be taken into account is that this number of clusters chosen is consistent with published works (e.g., \citet{2013ApJ...768...14W} and \citet{2017AA...604A..83B}), where two regimes were introduced in the planetary mass-radius relation by different techniques.

As for DBSCAN and OPTICS algorithms, they give $K$=2 but cannot separate clusters well; thus, we exclude them from the analysis. The effectiveness of clustering methods depends on several factors, including the dataset's characteristics and underlying distribution. Different clustering algorithms make certain assumptions about the data structure and employ distinct approaches to identify clusters. Accordingly, their performance can vary based on how well these assumptions align with the dataset's properties. In cases where clustering algorithms fail to find appropriate clusters, there may be several potential explanations. One important factor is the distribution of the exoplanet data, which might not conform to the assumptions made by certain clustering algorithms. DBSCAN and OPTICS are two density-based methods that execute clustering by finding areas where data points are concentrated. They can discover arbitrarily shaped clusters, including non-spherical ones; they, however, might fail when the dataset is too sparse, and the density varies across the data, like in the case of exoplanet data \citep{moreira2005density,ahmad2015performance}.

\begin{table*}
  \centering
  \caption{Clustering algorithms, breakpoints of radius ($B_{Radius}$) and mass ($B_{Mass}$) in logarithmic space, number of planets in the first ($N_{1}$) and second ($N_{2}$) clusters along with adjusted hyperparameters introduced in the Scikit-learn library \citep{scikit-learn}. The one-dimensional Gaussian distribution is used to find the intersection point and introduce the relevant breakpoints. Default values are set for hyperparameters that are not listed. The Elbow method and Silhouette score have been used to find the optimal number of clusters equal to 2. DBSCAN and OPTICS algorithms fail to provide appropriate clusters.}
  \begin{tabular}{llllll}
  \hline
  Algorithm & B$_{Radius}$ & B$_{Mass}$ & N$_{1}$ & N$_{2}$ & Adjusted parameter \\
  \hline
  Affinity Propagation & 0.93     & 1.73     & 254     & 440     & damping=0.9, preference=-60 \\
  BIRCH & 0.90   & 1.72  & 247   & 447   & n\_clusters=2, threshold=0.01 \\
  DBSCAN & --     & --     & --     & --     & eps=0.2, min\_samples=25 \\
  Gaussian Mixture Model & 0.95  & 1.80   & 271   & 423   & n\_components=2 \\
  Hierarchical Clustering & 0.90   & 1.72  & 247   & 447   & n\_clusters=2 \\
  K-Means & 0.92  & 1.72  & 252   & 442   & n\_clusters=2 \\
  Mean Shift & 0.93    & 1.73     & 257     & 437     & bandwidth=0.9 \\
  Mini-Batch K-Means & 0.92  & 1.72  & 252   & 442   & n\_clusters=2 \\
  OPTICS & --     & --     & --     & --     & min\_samples=40 \\
  Spectral Clustering & 0.89  & 1.64  & 239   & 455   & n\_clusters=2 \\
  \hline
  \end{tabular}
  \label{tab1}
\end{table*}

\begin{table}
\centering
\caption{Breakpoints of mass ($B_{Mass}$) and radius ($B_{Radius}$) derived by previous studies and in this work.}
\label{tab2}
\begin{tabular}{lll}
\hline
Study & B$_{Mass}$ (M$_{\oplus}$) & B$_{Radius}$ (R$_{\oplus}$) \\
\hline
\citet{2013ApJ...768...14W} & 150   & -- \\
\citet{2015ApJ...810L..25H} & 95    & -- \\
\citet{2017ApJ...834...17C} & $130\pm{22}$ & -- \\
\citet{2017AA...604A..83B} & $124\pm{7}$ & $12.1\pm{0.5}$ \\
This work & 52.48 & 8.13 \\
\hline
\end{tabular}
\end{table}

\begin{figure*}
	\includegraphics[width=0.85\paperwidth]{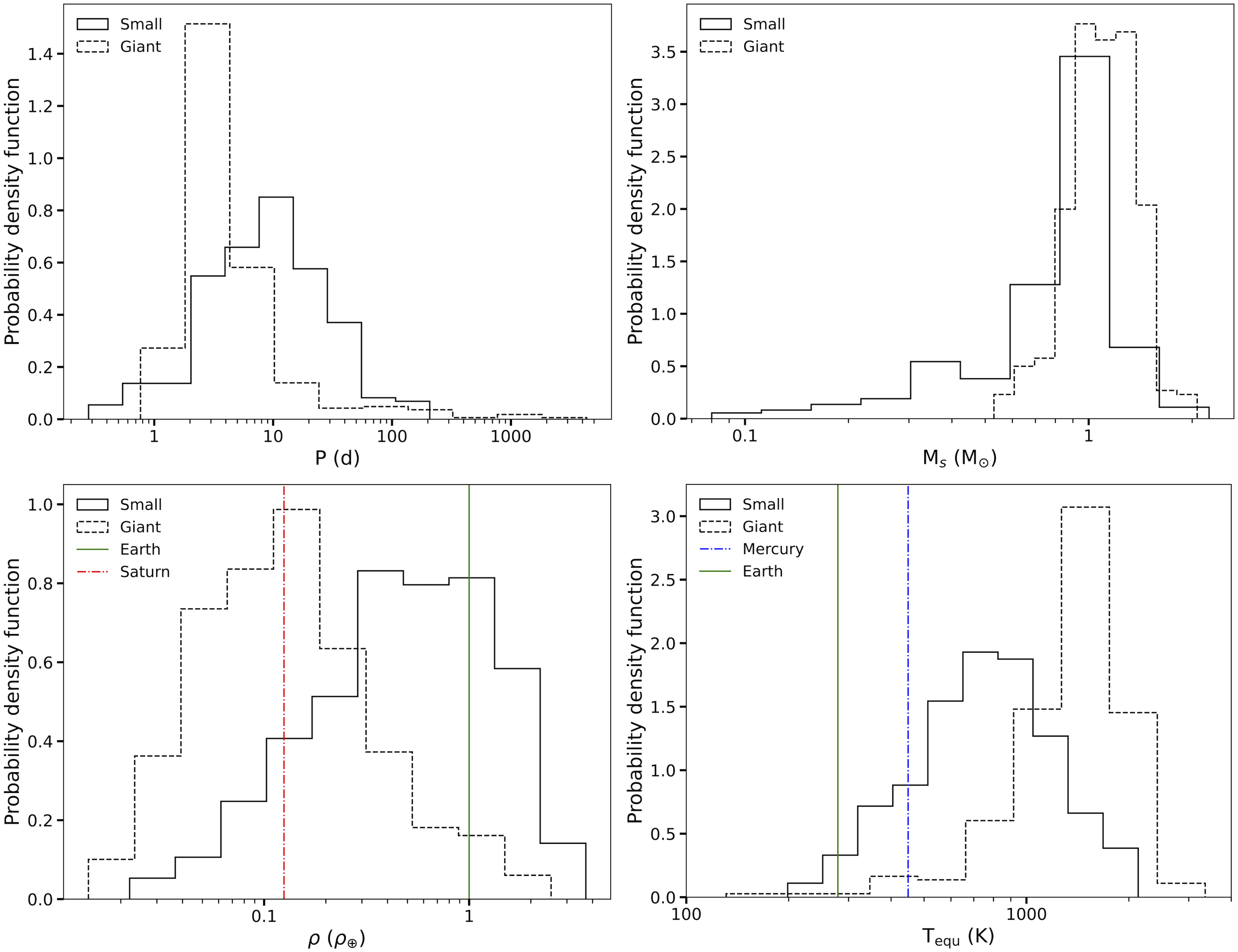}
    \caption{Histograms of the orbital period (upper-left panel), stellar mass (upper-right panel), average density (lower-left panel), and equilibrium temperature (lower-right panel) for small (solid-border bars) and giant (dashed-border bars) planets. The red dash-dotted line in the lower-left panel represents Saturn as the least dense planet, and the green solid line is Earth as the densest planet in the Solar System. In the lower-right panel, the green solid line represents Earth's equilibrium temperature, and the blue dash-dotted line is Mercury, which has the highest equilibrium temperature in the Solar System. The orbital period distribution shows that giant planets are closer to their host star than small planets. The stellar mass distribution demonstrates that, for most planets, the host star's mass is around the Sun's mass. Comparing planets' average density and calculated equilibrium temperature demonstrate that giant planets are hotter and less dense than small planets.}
    \label{fig9}
\end{figure*}

\subsubsection{Planet classes}
The planets are divided into two groups by choosing $K=2$ for BIRCH, Gaussian Mixture Model, Hierarchical Clustering, K-Means, Mini-Batch K-Means, and Spectral Clustering. Moreover, the Affinity Propagation and Mean Shift algorithms, which learn the number of clusters from data, result in two clusters. To introduce a boundary between two groups in the planet’s mass-radius space, we construct a Gaussian kernel density estimation for each cluster and find the intersection point. Table~\ref{tab1} lists the results of the clustering algorithms, which are almost similar (except for DBSCAN and OPTICS, as discussed in Sect.~\ref{number}). Ultimately, we separate data into two classes using an average value of $\log R_{p}=0.91$ ($R_{p}=8.13R_{\oplus}$) for radius breakpoint ($B_{Radius}$) and $\log M_{p}=1.72$ ($M_{p}=52.48M_{\oplus}$) for mass breakpoint ($B_{Mass}$). Exoplanets with $R_{p}\leq8.13R_{\oplus}$ and $M_{p}\leq52.48M_{\oplus}$ are defined as small planets, and those with $R_{p}>8.13R_{\oplus}$ and $M_{p}>52.48M_{\oplus}$ as giant planets. Fig.~\ref{fig8} shows the mass-radius distribution of clustered and outlier data. For several planets that lie outside the boundaries ($B_{Radius}$ and $B_{Mass}$), we use the criterion of their closeness to the boundaries to assign them to either of the classes.

According to a traditional definition, small exoplanets are planets with radii smaller than $4R_{\oplus}$ and masses lower than $\sim30M_{\oplus}$ \citep{2010Sci...330..653H,2014PNAS..11112655M,2014ApJ...783L...6W}; however, this customary definition of small and large planets does not exactly match previous studies that have investigated a transition point in the mass-radius distribution of exoplanets. Table~\ref{tab2} compares our breakpoints with those found by others in the literature and shows a considerable difference between mass breakpoints. Besides an increasing number of exoplanets and the evolution of their mass-radius distribution, this difference could result from applying different methods to find a cut-off point in planetary masses. Performed methods vary from a simple visual investigation of mass-radius and mass-density distributions \citep{2013ApJ...768...14W} to using different slope criteria in the planetary parametric relations \citep{2015ApJ...810L..25H,2017ApJ...834...17C,2017AA...604A..83B}. As a result, the mass and radius breakpoints identified in this work ($52.48M_{\oplus}$ and $8.13R_{\oplus}$, respectively) are closer to traditional breakpoints than those found in previous studies.

There are 254 small planets and 440 giant planets. The distributions of the orbital period, stellar mass, average density, and calculated equilibrium temperature for small and giant planets are demonstrated in Fig.~\ref{fig9}. The upper-left panel demonstrates that most giant planets are closer to their host star than small planets are. They, nonetheless, have a $P$ varying from 0.77 to 4331.01 days, while for small planets, it is between 0.28 and 207.62 days. The upper-right panel shows that the host star mass is frequently in a range around the Solar mass for most planets. It results from a selection effect: exoplanet-search programs often concentrate on Sun-like stars. In addition to this, lower-mass stars are not as likely to host exoplanets with sufficient mass to be identified by the radial-velocity technique \citep{2005A&A...443L..15B,2008PASP..120..531C}. Small planets revolve around stars with $M_{s}$ between 0.08 and 2.24 $M_{\odot}$, whilst, for host stars of giant planets, they vary from 0.53 to 2.07 $M_{\odot}$.

The lower-left panel compares small and giant planets' average density distribution. As expected, giant planets are generally less dense than small planets. The density distribution of giant planets peaks at about Saturn’s density. Hence, a significant fraction of giant planets has an average density similar to Saturn, the least dense planet in the Solar system. On the contrary, small planets cover a higher density range that includes Earth, the densest planet in the Solar system. This, in turn, suggests that giant planets are composed mainly of hydrogen and helium envelopes, whereas heavier elements dominate small planets.

Comparing the calculated equilibrium temperature of planets (lower-right panel) demonstrates that giant planets are hotter than small planets. This temperature difference between the two planet classes is expected because giant planets are closer to their host star (as shown in the upper-left panel); in addition to this, the host stars are almost of the same spectral type (equivalently, stellar mass, as shown in the upper-right panel).

It is important to note that our sample of giant planets is dominated by gas giant exoplanets with orbital periods of less than 10 days, commonly referred to as hot Jupiters. It is believed that these planets are more likely to be found around metal-rich stars than around stars with low stellar metallicity \citep{2018A&A...612A..93M,2020MNRAS.491.4481O,2023ApJ...949L..21Y}.

\begin{table*}
\centering
\caption{Results obtained by different regression algorithms. These algorithms have been separately applied to entire, small, and giant planets. Column 1 presents the name of the algorithm. Columns 2, 3, and 4 present the root mean square error (RMSE), mean absolute error (MAE), and coefficient of determination ($\rho^{2}$) for the whole dataset. Columns 5 to 10 present RMSE, MAE, and $\rho^{2}$ for small and giant planets, respectively. The last column lists adjusted hyperparameters introduced in the Weka environment \citep{witten2005practical,hall2009weka}.}
\label{tab3}
\begin{tabular}{llllllllllp{13.835em}}
\hline
Algorithm & RMSE  & MAE   & $\rho^{2}$ & RMSE$_{1}$ & MAE$_{1}$ & $\rho^{2}_{1}$ & RMSE$_{2}$ & MAE$_{2}$ & $\rho^{2}_{2}$ & \multicolumn{1}{l}{Adjusted parameter} \\
\hline
Bootstrap Aggregation & 0.096 & 0.068 & 0.933 & 0.126 & 0.096 & 0.692 & 0.064 & 0.047 & 0.489 & numIterations=100\newline{}classifier=REPTree \\
Decision Tree & 0.108 & 0.077 & 0.916 & 0.142 & 0.109 & 0.616 & 0.071 & 0.051 & 0.392 & maxDepth=-1\newline{}minNum=2 \\
K-Nearest Neighbors & 0.104 & 0.075 & 0.921 & 0.140 & 0.106 & 0.627 & 0.072 & 0.053 & 0.390 & KNN=3\newline{}distanceFunction=Euclidean \\
Linear Regression & 0.155 & 0.127 & 0.822 & 0.128 & 0.099 & 0.683 & 0.069 & 0.051 & 0.402 & eliminateColinearAttributes=True\newline{}attributeSelectionMethod=M5 \\
Multilayer Perceptron & 0.112 & 0.084 & 0.909 & 0.131 & 0.103 & 0.670 & 0.064 & 0.048 & 0.491 & learningRate=0.1\newline{}decay=False\newline{}momentum=0.1\newline{}hiddenLayers=a \\
M5P   & 0.098 & 0.071 & 0.930 & 0.130 & 0.101 & 0.688 & 0.073 & 0.053 & 0.391 & unpruned=False \\
Random Forest & 0.097 & 0.068 & 0.932 & 0.128 & 0.096 & 0.682 & 0.064 & 0.047 & 0.485 & numIterations=100\newline{}maxDepth=0\newline{}numFeatures=2 \\
SVR   & 0.093 & 0.065 & 0.937 & 0.123 & 0.088 & 0.710 & 0.063 & 0.046 & 0.510 & c=1\newline{}kernel=Puk \\
\hline
\end{tabular}
\end{table*}

\begin{table*}
\centering
\caption{Parametric equations obtained by different regression algorithms. Rows 1 to 7 present a linear fit, which equates the logarithm of planetary radius ($R_{p}$) to logarithms of planetary mass ($M_{p}$), orbital period ($P$), and stellar mass ($M_{s}$) plus a constant term ($C$) (see Eq.~\ref{eq5}). Row 8 presents a linear fit between planetary mass and radius as $\log(R_{p}/R_{\oplus})=A_{M_{p}}\log(M_{p}/M_{\oplus})+C$. The linear fit between the planetary radius and stellar mass is also presented in row 9 as $\log(R_{p}/R_{\oplus})=A_{M_{s}}\log(M_{s}/M_{\odot})+C$. The first column shows the row number. Column 2 presents the used dataset: the entire dataset, small planets, or giant planets. Best-fit parameters are listed in columns 3 to 6. The last column presents the applied algorithm: Linear Regression, M5P, or MCMC.}
\label{tab4}
\begin{tabular}{lllllll}
\hline
\# & Dataset & A$_{M_{p}}$ & A$_{P}$ & A$_{M_{s}}$ & C     & Algorithm \\
\hline
1     & Entire & 0.367 & -0.030 & 0.280 & 0.191 & Linear Regression \\
2     & Small & 0.467 & 0.090 & 0     & -0.103 & Linear Regression \\
3     & Giant & 0     & -0.069 & 0.480 & 1.157 & Linear Regression \\
4     & Small & 0.481 & 0.076 & 0.016 & -0.095 & M5P \\
5     & Giant & 0.012 & -0.067 & 0.489 & 1.123 & M5P \\
6     & Small & $0.482\substack{+0.025\\-0.024}$ & $0.078\substack{+0.017\\-0.016}$ & $0.031\substack{+0.041\\-0.042}$ & -$0.099\substack{+0.030\\-0.030}$ & M5P and MCMC \\
7     & Giant & $0.013\substack{+0.010\\-0.009}$ & -$0.070\substack{+0.007\\-0.007}$ & $0.492\substack{+0.036\\-0.036}$ & $1.121\substack{+0.024\\-0.024}$ & M5P and MCMC \\
8     & Small & $0.497\substack{+0.023\\-0.022}$ & --     & --     & -$0.050\substack{+0.024\\-0.024}$ & MCMC \\
9     & Giant & --     & --     & $0.480\substack{+0.036\\-0.037}$ & $1.109\substack{+0.004\\-0.004}$ & MCMC \\
\hline
\end{tabular}
\end{table*}

\begin{figure}
	\includegraphics[width=1.0\columnwidth]{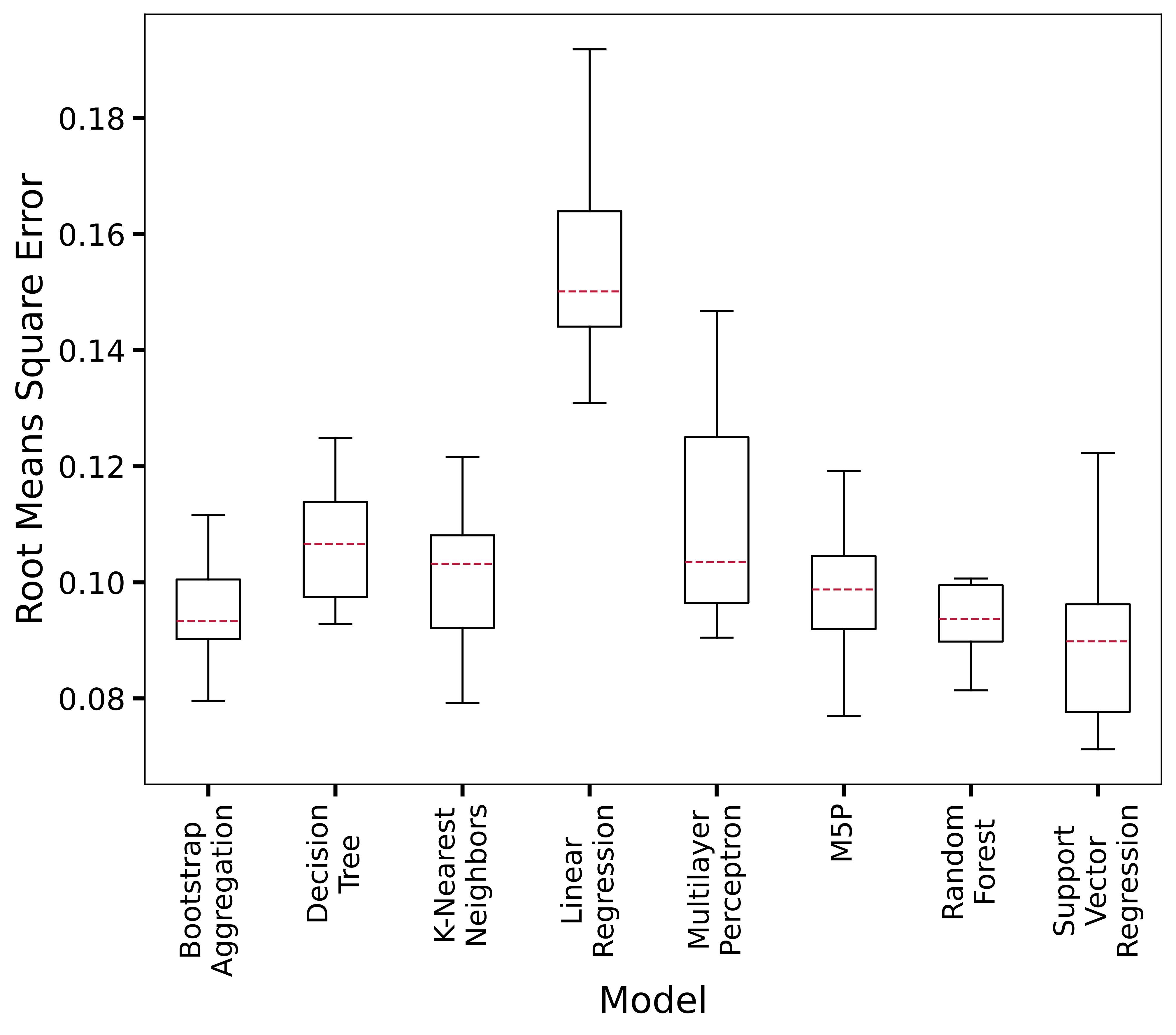}
    \caption{Box plots showing the spread of accuracy in the 10-fold cross-validation for predictive algorithms. On each box, the red dashed mark is the median, and the edges of the box are the $25^{th}$ and $75^{th}$ percentiles. Support Vector Regression with an RMSE of 0.093 is the best-performing model for the entire dataset, followed by Bootstrap Aggregation, Random Forest, and M5P with RMSEs of 0.096, 0.097, and 0.098, respectively.}
    \label{fig10}
\end{figure}

\begin{figure}
	\includegraphics[width=1.0\columnwidth]{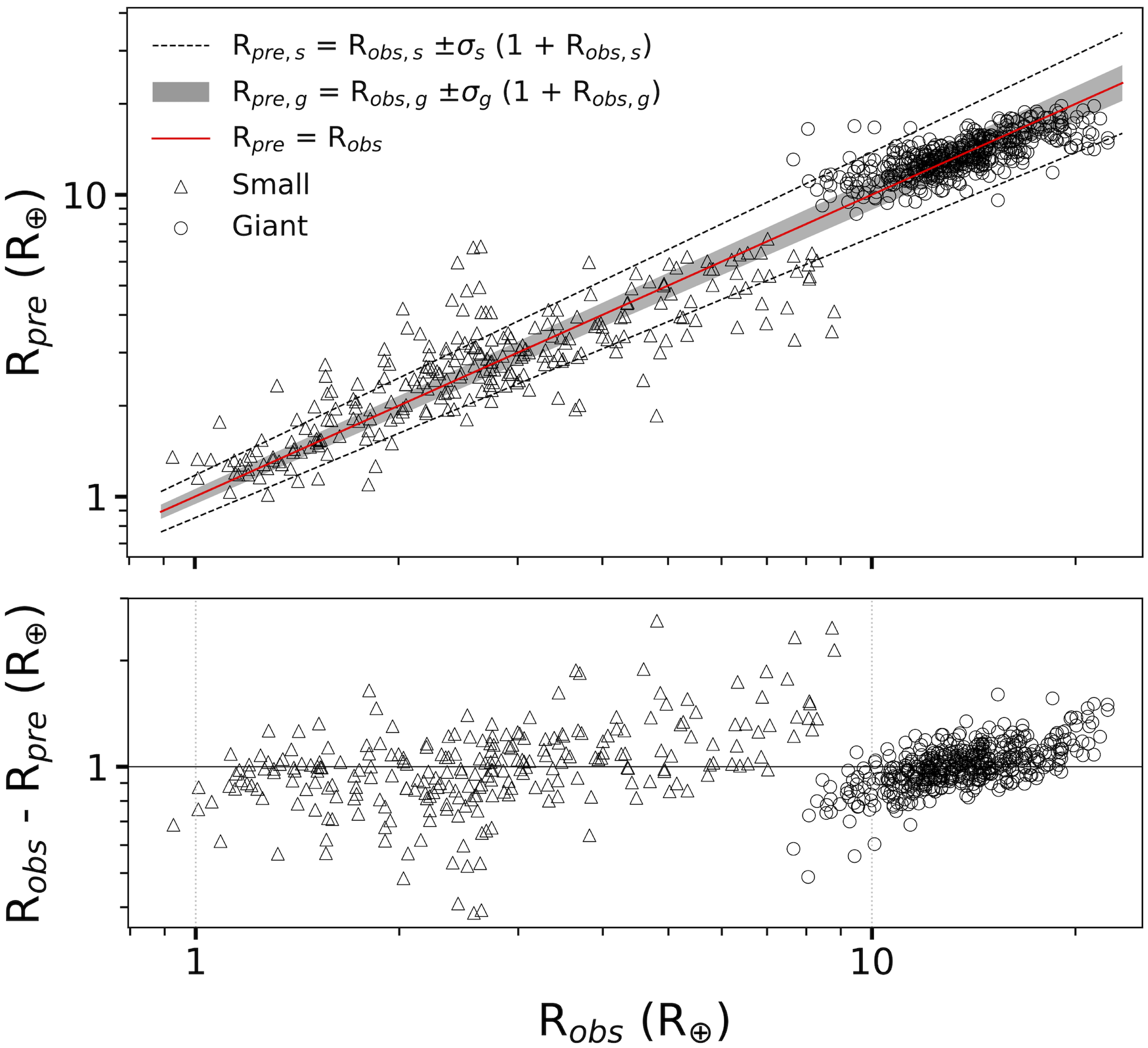}
    \caption{Comparison between observed ($R_{obs}$) and predicted ($R_{pre}$) planetary radius (upper panel) along with residual values (lower panel) obtained by the SVR model, which has been applied separately to small (triangles) and giant (circles) planets. The red line indicates $R_{pre}=R_{obs}$ along with two dashed lines and a gray area that illustrate the normalized median absolute deviation (NMAD) for small and giant planets, respectively. Considering \citet{1983ured.book.....H}, NMAD is calculated using $R_{pre}=R_{obs}\pm\sigma(1+R_{obs})$, where $\sigma=1.48\times median[\mid R_{pre}-R_{obs} \mid/(1+R_{obs})]$.}
    \label{fig11}
\end{figure}

\begin{figure}
\centering
	\includegraphics[width=1.0\columnwidth]{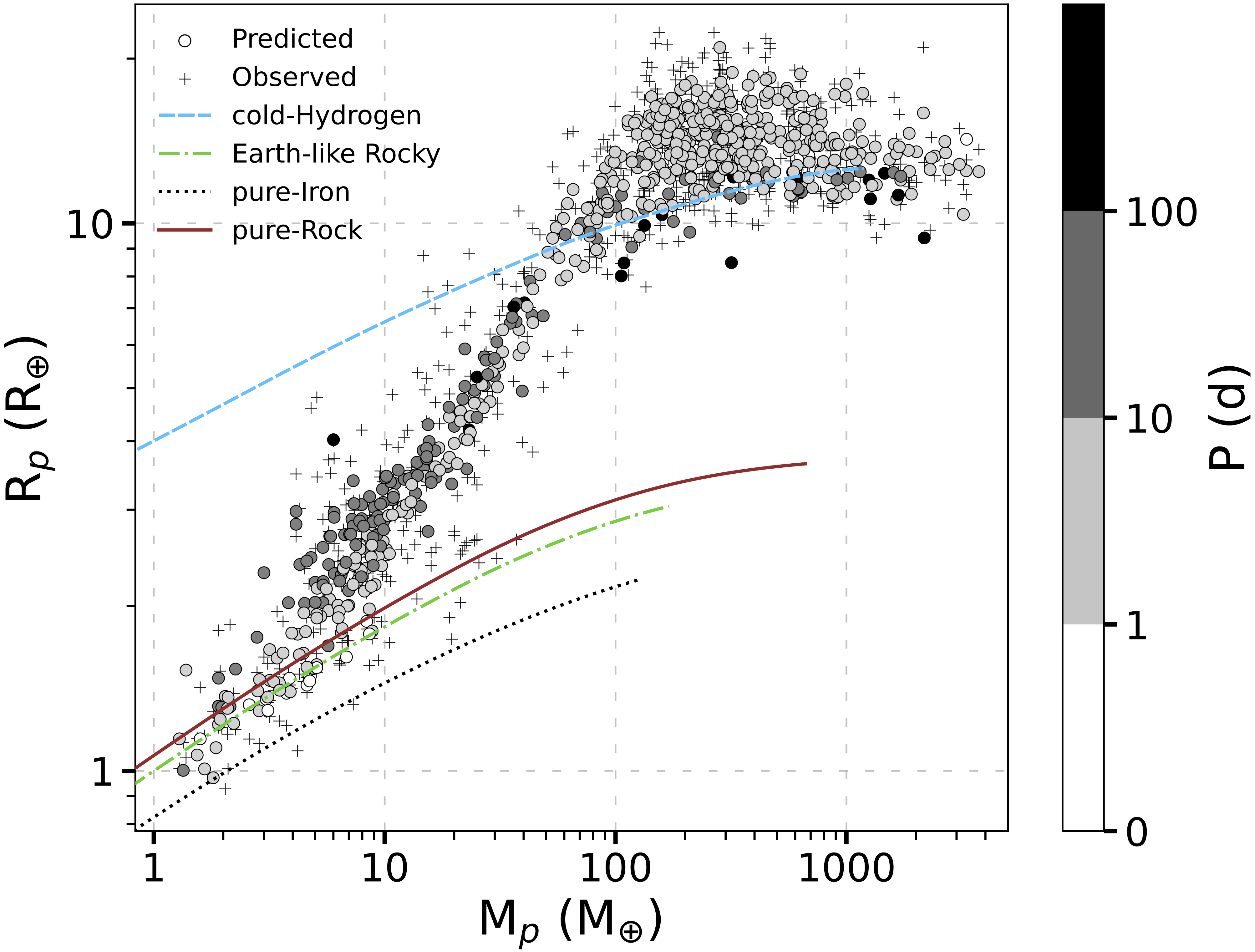}
    \caption{Predicted (circles) and observed (pluses) radius as a function of a mass and orbital period obtained by the SVR model for all planets in the sample. The distribution of cold-hydrogen (blue dashed line), Earth-like rocky (green dash-dotted line), pure-iron (black dotted line), and pure rocky (solid crimson line) planets are also illustrated \citep{2010ApJ...712L..73M,2014ApJS..215...21B}.}
\label{fig12}
\end{figure}

\subsection{Prediction of the planetary radius}
To find the radius of a planet based on physical parameters, ML predictive algorithms are applied to our dataset of 694 planets. The physical parameters are the planet's mass ($M_{p}$), orbital period ($P$), and host star's mass ($M_{s}$), selected by FS methods. Bootstrap Aggregation, Decision Tree, K-Nearest Neighbors, Linear Regression, Multilayer Perceptron, M5P, Random Forest, and Support Vector Regression (SVR) are implemented algorithms. These algorithms are applied separately to entire, small, and giant planets. To have the best performance of algorithms, the hyperparameters are tuned. Furthermore, a 10-fold cross-validation procedure is used to assess the performance of models. The root means square error (RMSE), mean absolute error (MAE), and coefficient of determination ($\rho^{2}$) are calculated as validation metrics (see Eq.~\ref{eq2},~\ref{eq3}, and~\ref{eq4}).

RMSE, MAE, and $\rho^{2}$ of entire, small, and giant planets, along with the tuned hyperparameters, are listed in Table~\ref{tab3}. Fig.~\ref{fig10} shows the box plots of accuracy in the 10-fold cross-validation for models. SVR with an RMSE of 0.093 is the best-performing model for the entire dataset, followed by Bootstrap Aggregation, Random Forest, and M5P with RMSEs of 0.096, 0.097, and 0.098, respectively. They also have lower MAE and higher $\rho^{2}$ values than other models. The SVR performs better than other algorithms for both subsets of small and giant planets.

Fig.~\ref{fig11} compares the observed ($R_{obs}$) and predicted ($R_{pre}$) radius (upper panel) along with residual values (lower panel) obtained by SVR as the best-performing model. In this figure, the model has been applied separately to small and giant planets, resulting in a gap and a relatively higher dispersion around $\sim8R_{\oplus}$. $\rho^{2}$ value is 0.710 for small planets and 0.510 for giant planets. The normalized median absolute deviation (NMAD) has been calculated for small and giant planets predictions. NMAD is $R_{pre}=R_{obs}\pm\sigma(1+R_{obs})$, where $\sigma=1.48\times median[\mid R_{pre}-R_{obs} \mid/(1+R_{obs})]$ \citep{1983ured.book.....H}. As observed in the lower panel of Fig.~\ref{fig11}, a distinct linear trend with a slope of 0.459 is noticeable in the residuals between predicted and observed radii of giant planets. This trend could be attributed to factors such as systematic observation errors, the determination of physical parameters, and calculation issues including hyperparameter tuning, algorithm characteristics and limitations, and selected features. We adjust the learning algorithms to address this challenge by re-configuring their hyperparameters. Interestingly, a similar pattern emerges in the residual values across all eight predictive algorithms. This suggests that the constraints of SVR alone cannot explain this pattern. Moreover, we discover that excluding or including the parameter $M_{p}$ significantly affects the slope of the linear trend ($\pm0.08$) compared to other parameters. However, modifying the subset of features does not eliminate this trend. The most gradual trend appears when utilizing the feature subset of $M_{p}$, $P$, and $M_{s}$.

It is important to note that this trend minimally impacts radius prediction. Additionally, the residuals between predicted and observed radii of giant planets fall within the range seen for small planets. It is plausible that this trend is linked to systematic issues in exoplanet observations or the determination of observed parameters (e.g., mass, orbital period, and radius). As the current sample of exoplanets detected through the transit method is substantial, further investigation of this effect could be undertaken when a statistically significant sample of exoplanets detected through other detection methods becomes available.

Fig.~\ref{fig12} shows the predicted and observed radii as a function of a mass and orbital period obtained by the SVR model. The model has been applied to the entire sample in this figure, resulting in an $\rho^{2}$ of 0.937. SVR can efficiently reproduce the spread in radius.

\begin{figure}
\centering
	\includegraphics[width=1.0\columnwidth]{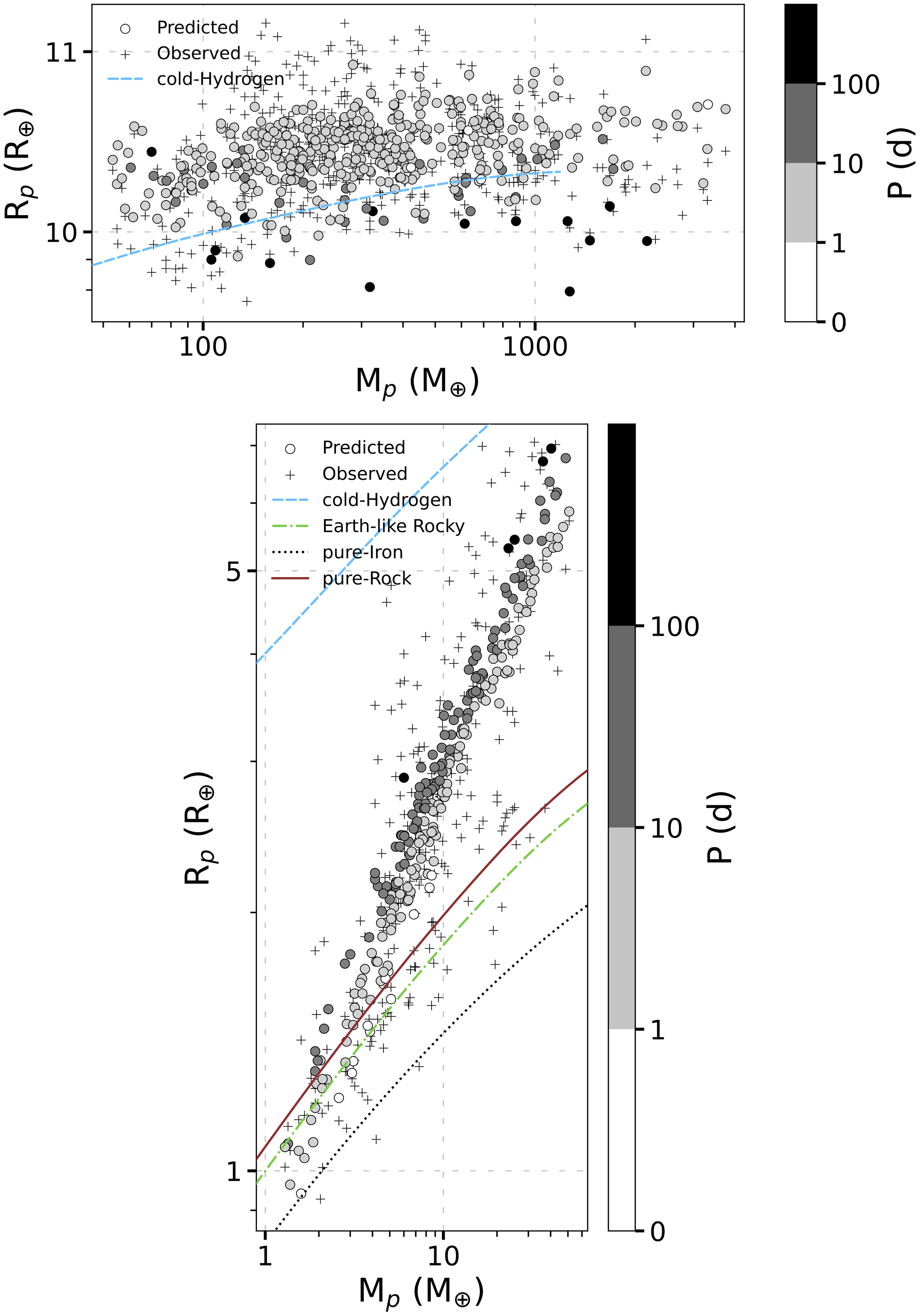}
    \caption{Predicted (circles) and observed (pluses) radius as a function of a mass and orbital period obtained by the M5P model for small (lower panel) and giant (upper panel) planets. Four iso-density curves are also drawn: cold-hydrogen (blue dashed line), Earth-like rocky (green dash-dotted line), pure-iron (black dotted line), and pure rocky (solid crimson line) planets \citep{2010ApJ...712L..73M,2014ApJS..215...21B}.}
\label{fig13}
\end{figure}

Linear Regression and M5P can derive parametric equations between physical parameters by fitting linear models to the exoplanet data. A linear model is fitted to the real data in the Linear Regression algorithm. At the same time, M5P splits the entire dataset into several subsets and fits a multivariate linear function to each subset. Eq.~\ref{eq5} presents a linear fit between the planetary radius, planetary mass, orbital period, and stellar mass derived by Linear Regression and M5P, where $A_{M_{p}}$, $A_{P}$, $A_{M_{s}}$, and $C$ are fitting parameters.

\begin{equation}
\log \biggl(\frac{R_{p}}{R_{\oplus}}\biggr)=A_{M_{p}}\log \biggl(\frac{M_{p}}{M_{\oplus}}\biggr)+A_{P}\log \biggl(\frac{P}{d}\biggr)+A_{M_{s}}\log \biggl(\frac{M_{s}}{M_{\odot}}\biggr)+C.
\label{eq5}
\end{equation}

Best-fit parameters obtained by Linear Regression and M5P are listed in Table~\ref{tab4}. Row 1 presents the linear fit of all planets provided by the Linear Regression algorithm. Running this algorithm independently for clusters produces individual linear fit for each cluster (rows 2 and 3). For small planets $A_{M_{s}}=0$, and for giant planets $A_{M_{p}}=0$, implying no dependence between the planetary radius and stellar mass of small planets, as well as between the radius and mass of giant planets.

M5P divides the planets into two groups using a mass breakpoint of $\log M_{p}=1.717$ ($M_{p}=52.12M_{\oplus}$). Interestingly, clustering algorithms also find this breakpoint (see Table~\ref{tab1}). Rows 4 and 5 present multivariate linear fits of small and giant planets produced by the M5P algorithm. Splitting the data provides M5P with much better results than Linear Regression (see Table~\ref{tab3}). To estimate the uncertainty values of the M5P's best-fit parameters, we use the MCMC method. The likelihood function and initial values implemented in the MCMC analysis are the same as those acquired by the M5P. In Table~\ref{tab4}, rows 6 and 7 present the linear fits of small and giant planets, respectively, together with the uncertainty values obtained by the MCMC method. Fig.~\ref{fig13} presents the distributions of the predicted and observed radii as a function of mass and orbital period, obtained by the M5P model for small (lower panel) and giant (upper panel) planets. The value of $\rho^{2}$ for the whole sample is 0.930. The predicted radii reproduce the spread in radius, especially for giant planets.

\subsection{Dependence of planetary radius on host star's mass}
There are inconsistent assertions in the literature about the dependence of planetary parameters on the host star's mass. \citet{2018ApJ...856L..28P} claimed that the mass of the most common exoplanets depends on their host star mass. They investigated G, K, and M stars and suggested that planets around relatively low-mass stars (with a mass lower than $1M_{\odot}$) are lower in mass and smaller in radius. In contrast, \citet{2018ApJ...858...58N} showed that the mass-radius relation of small planets has no strong dependence on stellar mass. \citet{2019ApJ...874...91W} discussed a linear relationship between exoplanet mass and host star mass and the lack of correlation between the planetary radius and stellar metallicity. In addition, \citet{2021A&A...652A.110L} studied exoplanets with radii up to $8R_{\oplus}$ and masses up to $20M_{\oplus}$ surrounding G and K stars. They confirmed that exoplanets revolving around more massive stars tend to be larger and more massive.

\begin{figure*}
\centering
	\includegraphics[width=0.85\paperwidth]{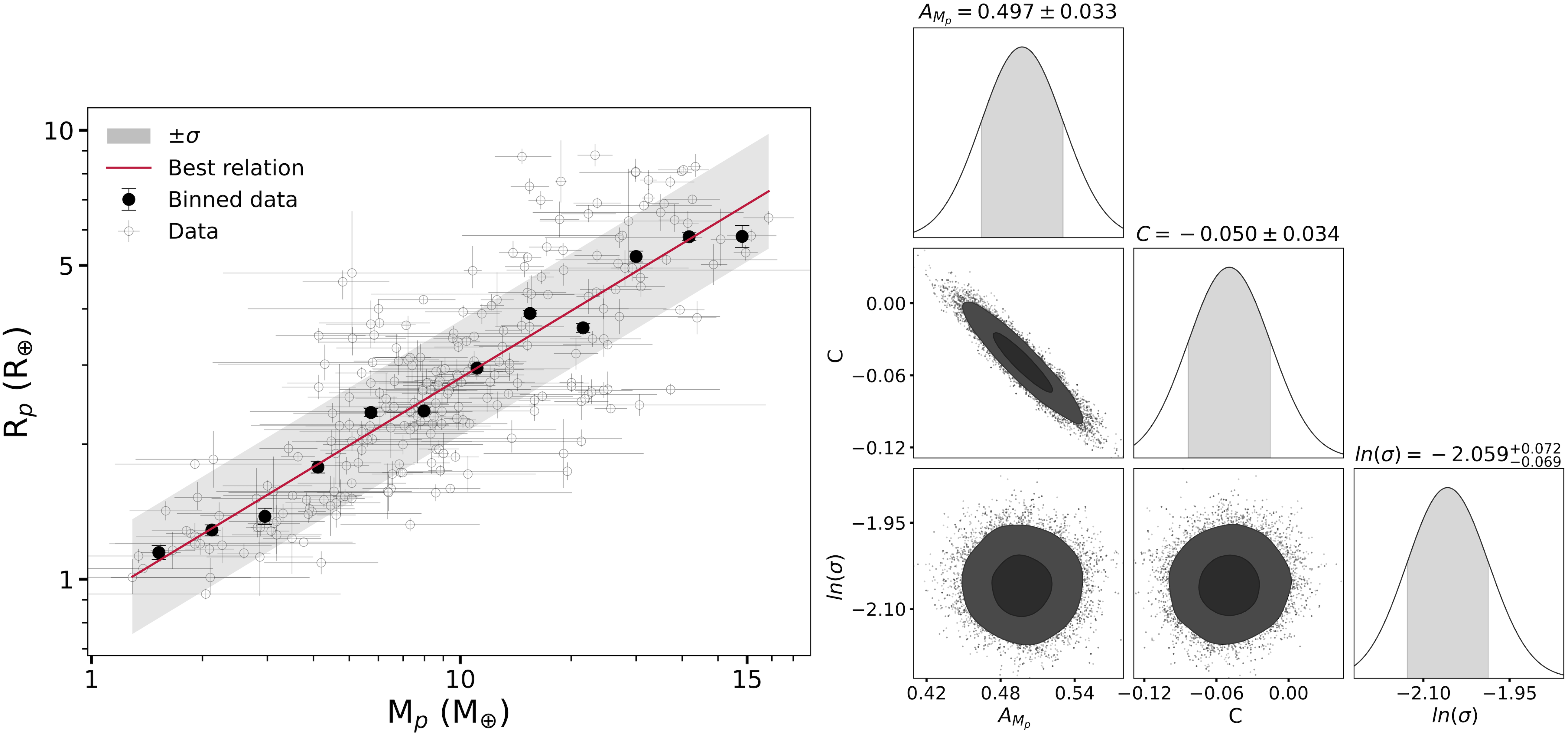}
    \caption{Left panel: the relation between mass and radius of small planets obtained by the MCMC method. The red line is the best scaling relation plotted using $\log(R_{p}/R_{\oplus})=0.497\log(M_{p}/M_{\oplus})-0.050$ (see Table~\ref{tab4}, row 8). The gray area is $\pm1\sigma$ uncertainties around the best scaling relation. For better illustration, the data points have been binned with a width of 0.05 $\log M_{p}$. The binned and actual data are shown in black and gray circles. Right panel: the one- and two-dimensional marginalized posterior distributions of the scaling relation parameters obtained by the MCMC method. $A_{M_{p}}$ and $C$ are the slope and intercept, respectively. The uncertainty around the best scaling relation is shown by $\sigma$.}
    \label{fig14}
\end{figure*}

\begin{figure*}
\centering
	\includegraphics[width=0.85\paperwidth]{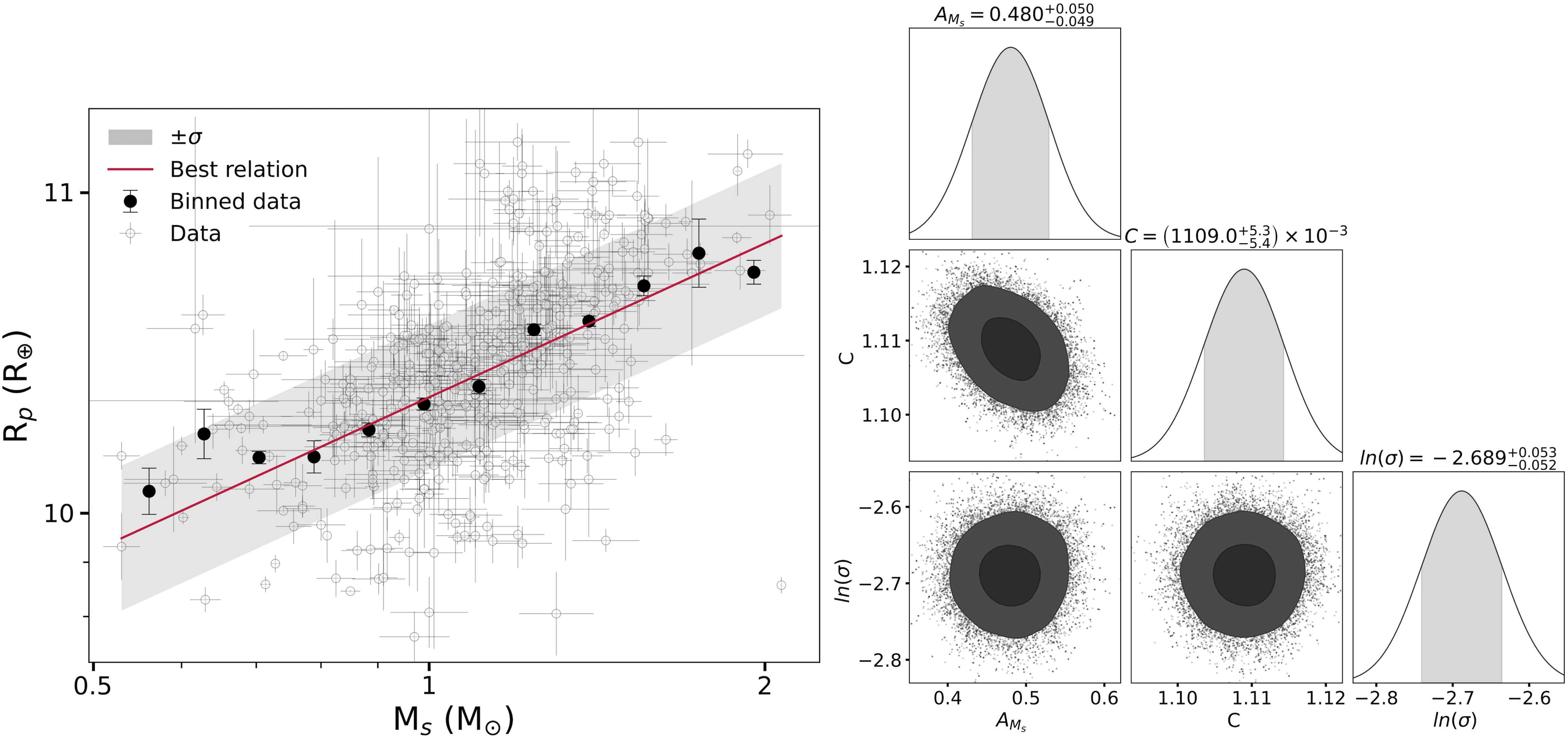}
    \caption{Left panel: the relation between the radius of giant planets and the host star's mass obtained by the MCMC method. The red line is the best scaling relation plotted using $\log(R_{p}/R_{\oplus})=0.480\log(M_{s}/M_{\odot})+1.109$ (see Table~\ref{tab4}, row 9). The gray area is $\pm1\sigma$ uncertainties around the best scaling relation. For better illustration, the data points have been binned with a width of 0.05 $\log M_{s}$. The binned and actual data are shown in black and gray circles. Right panel: the one- and two-dimensional marginalized posterior distributions of the scaling relation parameters obtained by the MCMC method. $A_{M_{s}}$ and $C$ are the slope and intercept, respectively. The uncertainty around the best scaling relation is shown by $\sigma$.}
\label{fig15}
\end{figure*}

As can be seen in Table~\ref{tab4}, the radius of a small planet shows a strong dependency on its mass. Furthermore, there is no strong correlation between stellar mass and planetary radius for small planets. In comparison, the radius of a giant planet depends weakly on its mass because above $\sim8R_{\oplus}$ the electron degeneracy pressure dominates \citep{1969ApJ...158..809Z,2007ApJ...669.1279S,2012ApJ...744...59S}. In addition to this, the planetary radius and stellar mass of giant planets have a strong linear correlation. We apply the MCMC as a supportive method to find the best scaling relations between the radius and mass of small planets and between the radius and stellar mass of giant planets while considering the reported errors of physical values. Row 8 of Table~\ref{tab4} presents the linear fit between the radius and mass of small planets. Additionally, the linear fit between the radius of giant planets and the mass of their host stars is presented in row 9. The related diagrams are depicted in Fig.~\ref{fig14} and \ref{fig15}.

Giant planets are less dense than small planets (see the lower-left panel of Fig.~\ref{fig9}) and mostly composed of volatile elements (hydrogen and helium envelopes). On the other hand, giant planets orbit stars more massive than $\sim1M_{\odot}$, whereas the hosts of small planets include low-mass stars; that is, they have a mass greater than $0.08M_{\odot}$ (see the upper-right panel of Fig.~\ref{fig9}). Concentrating on a limited sample of exoplanets, \citet{2021A&A...652A.110L} concluded that planets forming around massive stars accrete more H-He atmospheres than those that form around low-mass stars. Our extensive sample suggests a similar scenario for giant planets. Hence, the dependence of the radius of giant planets on the host star's mass may result from their different planetary composition. It should be noted that, in addition to naturally correlated parameters, this trend with stellar mass might be a consequence of observational biases. The larger transiting exoplanets are more detectable around luminous stars with larger masses \citep{2016arXiv160700322B}.

\subsection{Effect of equilibrium temperature, semi-major axis, and luminosity}
\begin{figure}
	\includegraphics[width=\columnwidth]{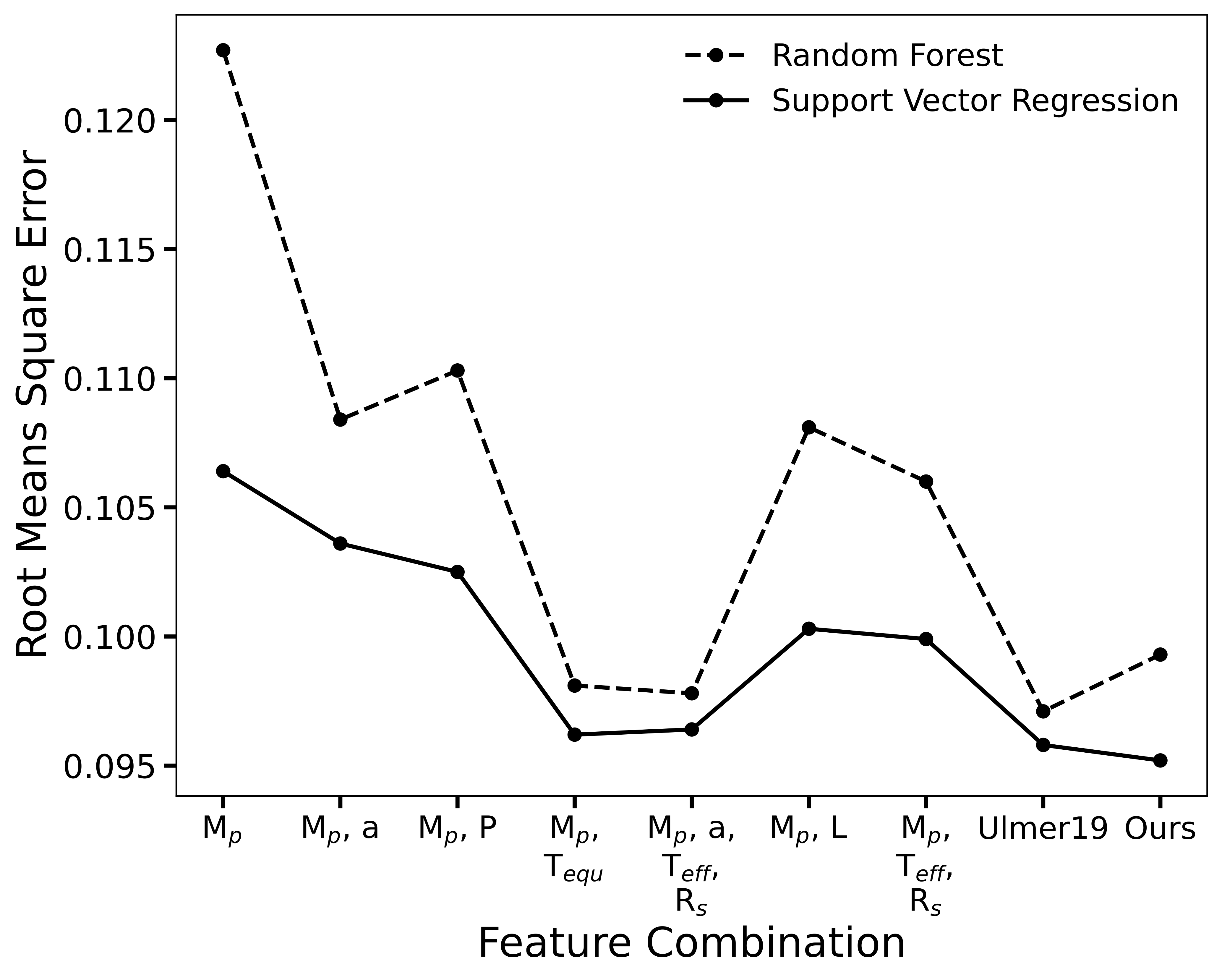}
    \caption{Comparison of the performance of Random Forest and Support Vector Regression (SVR) models for different feature combinations. The dataset consists of 506 planets collected by \citetalias{2019A&A...630A.135U} and transferred to a logarithmic space. The dashed and solid lines represent root means square errors obtained by Random Forest and SVR, respectively. Features are as follows: orbital period ($P$), planetary mass ($M_{p}$), semi-major axis ($a$), equilibrium temperature ($T_{\text{equ}}$), luminosity ($L$), stellar mass ($M_{s}$), stellar radius ($R_{s}$), and effective temperature ($T_{\text{eff}}$). The eighth set consists of $M_{p}$, $T_{\text{equ}}$, $a$, $R_{s}$, $M_{s}$, $L$, and $T_{\text{eff}}$, selected by \citetalias{2019A&A...630A.135U}. The last set is our feature combination, which contains $M_{p}$, $P$, and $M_{s}$. The SVR model performs better than the Random Forest model for all combinations.}
    \label{fig16}
\end{figure}

\citet{2016arXiv160700322B} used a Random Forest model to assess the effect of different physical parameters on predicting a planet's radius. They concluded that the planet's mass and equilibrium temperature have the greatest effect. In a similar work, \citetalias{2019A&A...630A.135U} presented planetary mass, equilibrium temperature, semi-major axis, stellar radius, mass, luminosity, and effective temperature as important parameters, and stellar metallicity, orbital period, and eccentricity as the three least important parameters. We add orbital periods to the dataset collected by \citetalias{2019A&A...630A.135U} and transfer it to a logarithmic space. Hence, a new dataset consisting of 506 planets is provided. The features of this dataset are as follows: orbital period ($P$), planetary mass ($M_{p}$), semi-major axis ($a$), equilibrium temperature ($T_{\text{equ}}$), luminosity ($L$), stellar mass ($M_{s}$), stellar radius ($R_{s}$), and effective temperature ($T_{\text{eff}}$). To evaluate the effect of equilibrium temperature, semi-major axis, and luminosity in predicting planetary radius and to compare the performance of Random Forest and SVR models in \citetalias{2019A&A...630A.135U}'s dataset, we implement models on the nine feature combinations. As the most important parameter, planetary mass is added to all combinations. Fig.~\ref{fig16} presents the RMSE values obtained by Random Forest and SVR versus different feature combinations. The SVR model performs better than Random Forest for all combinations.

According to Kepler's third law, the orbital period and semi-major axis are correlated to each other \citep{cox2015allen}. So, as expected, considering the SVR model, the RMSE values of the second (0.104) and third (0.103) combinations are not significantly different. The second combination consists of $M_{p}$ and $a$, while the third set includes $M_{p}$ and $P$. Moreover, the equilibrium temperature of a planet can be calculated using Eq.~\ref{eq6} without considering the effect of albedo and eccentricity \citep{2015trge.book..673L}.

\begin{equation}
T_{\text{equ}}=\sqrt{\frac{R_{s}}{2a}}\times T_{\text{eff}}.
\label{eq6}
\end{equation}

The fourth combination includes planetary mass and equilibrium temperature, and the fifth combination includes planetary mass and constituent parameters of equilibrium temperature ($a$, $R_{s}$, and $T_{\text{eff}}$). The SVR model's RMSE values corresponding to the fourth and fifth sets are almost identical (0.096).

The luminosity of a star is correlated to its radius and effective temperature ($L\propto R_{s}^{2}\times T_{\text{eff}}^{4}$). The sixth combination is planetary mass and luminosity, whose RMSE value (0.100) is almost the same as the seventh combination, which includes planetary mass and constituent parameters of luminosity ($T_{\text{eff}}$ and $R_{s}$). Additionally, the eighth set corresponds to features selected by \citetalias{2019A&A...630A.135U}, including $M_{p}$, $T_{\text{equ}}$, $a$, $R_{s}$, $M_{s}$, $L$, and $T_{\text{eff}}$, and the last set consists of $M_{p}$, $P$, and $M_{s}$, which we have selected. Using the SVR model, it is clear that there is no remarkable difference between the results obtained by these two feature sets. RMSE of our feature set equals 0.095 while 0.096 for that used by \citetalias{2019A&A...630A.135U}.

Although \citetalias{2019A&A...630A.135U} considered the orbital period as an inconsequential parameter, here we show that the orbital period (or semi-major axis), along with the planet's mass and one of the stellar parameters, have remarkable effects on predictions. Moreover, contrary to the results acquired by \citet{2016arXiv160700322B} and \citetalias{2019A&A...630A.135U}, it seems that considering stellar luminosity and planetary equilibrium temperature as features do not improve the accuracy of planetary radius predictions. Luminosity and equilibrium temperature are two physically dependent parameters. Thus, similar results can be achieved only by considering their constituent parameters.

\section{Summary and conclusions}\label{conclusion}
In this study, we conduct a comprehensive analysis of a sample comprising 762 exoplanets and eight Solar System planets. Our main objective is to investigate the characteristics of these exoplanets and explore the correlations between various features. The dataset includes essential parameters such as orbital period ($P$) and eccentricity ($e$), planetary mass ($M_{p}$), and radius ($R_{p}$), and the stellar mass ($M_{s}$), radius ($R_{s}$), metallicity (Fe/H), and effective temperature ($T_{\text{eff}}$).

To ensure the reliability of our analysis, we employ the Local Outlier Factor (LOF) algorithm, which allows us to identify and filter out data points that deviate significantly from the overall dataset. This process leads us to a refined dataset consisting of 76 anomalous objects, which can be considered as robust and reliable measurements for our subsequent analysis.

By utilizing Feature Selection (FS) methods, we determine the most influential factors in predicting the radius of exoplanets. Our findings highlight that planetary mass ($M_{p}$) plays a pivotal role in this regard, whereas eccentricity ($e$) and metallicity (Fe/H) demonstrate relatively lesser significance in the prediction process.

To further understand the underlying structure of the dataset, we employ various clustering algorithms and evaluation techniques such as the Elbow, Silhouette, and Hierarchical methods. Based on the outcomes of these analyses and in alignment with the conclusions drawn by \citet{2017AA...604A..83B}, we opt to divide the dataset into two distinct clusters: small and giant planets. Notably, we observe distinct breakpoints in the mass-radius space at $M_{p}=52.48M_{\oplus}$ and $R_{p}=8.13R_{\oplus}$ for these clusters.

Our analysis uncovers significant disparities between small and giant planets. Giant planets tend to exhibit higher masses, larger radii, and lower densities, suggesting a prevalence of volatile-rich exoplanets in this category. Additionally, these giant planets tend to orbit their host stars at closer distances and possess higher equilibrium temperatures. On the other hand, small planets predominantly consist of elements heavier than hydrogen and helium, exhibiting lower equilibrium temperatures.

To predict the planetary radius, we employ various Machine Learning (ML) regression models. Among these models, the Support Vector Regression (SVR) demonstrates superior performance, yielding a Root Mean Squared Error (RMSE) of $0.093$. A discernible linear trend appears in the residuals between predicted and observed radii of giant planets, which is not attributed to the restrictions of the predictive models or calculation issues. This evident trend has no significant impact on the radius predictions and is possibly related to the systematic issues in the exoplanet observations or the determination of physical parameters.

Additionally, utilizing Linear Regression, M5P, and Markov Chain Monte Carlo (MCMC) methods, we establish a positive linear mass-radius relationship for small planets. In contrast, the radius of giant planets exhibits a positive correlation with the mass of their host stars, consistent with the findings presented by \cite{2021A&A...652A.110L}, which suggest a connection between volatile-rich planets and more massive host stars. Nonetheless, as most of our sample consists of transiting exoplanets, besides naturally correlated parameters, the observational bias in the detection method can explain this result.

Furthermore, our analysis reveals that a carefully selected subset of features, encompassing planetary mass, orbital period, and one of the stellar parameters (stellar mass, radius, or effective temperature), is sufficient for accurate radius prediction. The inclusion of additional features such as semi-major axis, equilibrium temperature, and luminosity does not yield substantial improvements in the predictive capability.

Looking ahead, a comprehensive understanding of exoplanet composition and structure, as well as the testing of theories related to planetary formation and evolution, necessitates further follow-up observations.  The James Webb Space Telescope (JWST) and future missions such as the Extremely Large Telescope (ELT), the Atmospheric Remote-sensing Infrared Exoplanet Large-survey (ARIEL), and the Planetary Transits and Oscillations of Stars (PLATO) mission will undoubtedly contribute invaluable insights into the atmospheric characteristics of exoplanets and the determination of stellar ages, thereby facilitating a more detailed exploration of exoplanetary systems.

\section{Acknowledgements}
The authors thank the anonymous reviewers who helped with their valuable and priceless comments towards improving the manuscript. This research has made use of the NASA Exoplanet Archive, which is operated by the California Institute of Technology, under contract with the National Aeronautics and Space Administration under the Exoplanet Exploration Program. It has also made use of data obtained from or tools provided by the portal exoplanet.eu of The Extrasolar Planets Encyclopedia. Furthermore, this research has used the NASA’s Planetary Fact Sheet. The authors acknowledge the usage of the \texttt{Scikit-learn} library \citep{scikit-learn} and the \texttt{Weka} software \citet{witten2005practical,hall2009weka}. They also acknowledge the usage of the following python packages, in alphabetical order: \texttt{astropy} \citep{astro2013,astro2018}, \texttt{chainConsumer} \citep{2019ascl.soft10017H}, 
\texttt{emcee} \citep{emcee3}, \texttt{matplotlib} \citep{mathlib}, \texttt{numpy} \citep{numpy2}, and \texttt{scipy} \citep{scipy}.

\section*{Data Availability}
The data underlying this article were derived from sources in the public domain: NASA Exoplanet Archive (\url{https://exoplanetarchive.ipac.caltech.edu/}), Extrasolar Planets Encyclopedia (\url{http://exoplanet.eu/}), and NASA’s Planetary Fact Sheet (\url{https://nssdc.gsfc.nasa.gov/planetary/factsheet/}).



\bibliographystyle{mnras}
\bibliography{example} 




\appendix

\section{Data-cleaning and its impact on prediction accuracy}\label{appA}
Our dataset contains 770 data points. The Local Outlier Factor (LOF) method is chosen to identify outlier observations. It is first applied to all parameters including $P$, $e$, $M_{p}$, $R_{p}$, $M_{s}$, $R_{s}$, Fe/H, and $T_{\text{eff}}$, and then to $R_{p}$ and $M_{p}$. The first step determines 39 outliers with an average score of 1.951, where the higher the LOF score, the more abnormal the data point. In comparison, the average score of inliers is 1.113. The second step determines 37 outliers and assigns average scores of 2.113 and 1.061 to the outlier and inlier data points, respectively. In total, the LOF marks 76 data points as outliers. The outliers have an average Mahalanobis distance of 18.050, compared to 6.888 for the inliers. This indicates that the 76 outlier data points are farther away from the dataset's central point than the inliers. It is interesting to note that identified outliers have higher uncertainties than inliers. In a logarithmic scale, outlier data points have an average uncertainty of 0.19 for planetary mass and 0.05 for planetary radius. In comparison, these values for inlier data points are 0.11 and 0.04, respectively.

\begin{figure}
	\includegraphics[width=\columnwidth]{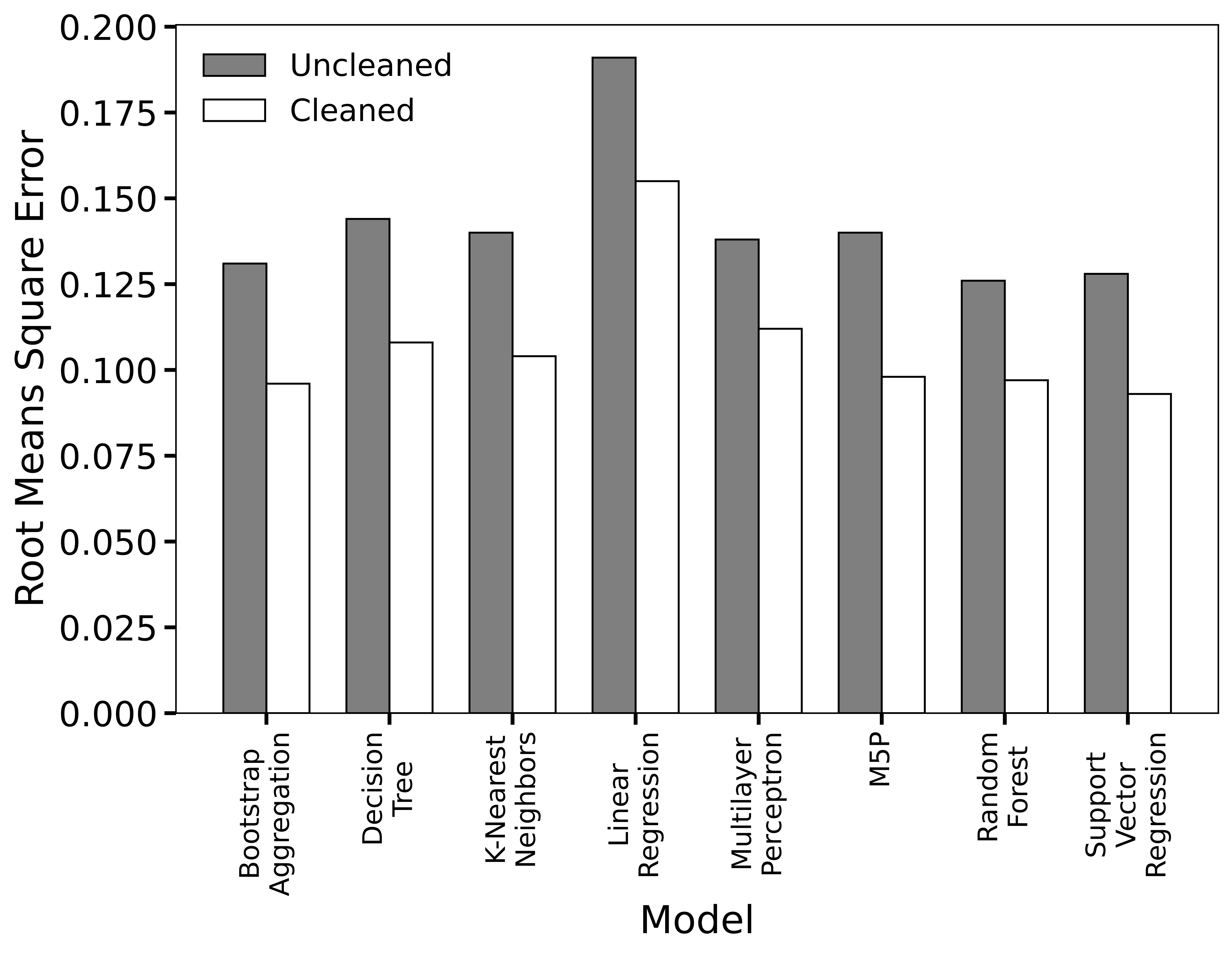}
    \caption{Root means square error (RMSE) values of different ML regression models implemented in uncleaned (gray bars) and cleaned (white bars) datasets. All models have higher RMSE values when outliers are included in the dataset.}
    \label{figA1}
\end{figure}

To quantify the impact of outliers on prediction precisions, we run ML regression models on the dataset containing all 770 data points. Fig.~\ref{figA1} compares the prediction accuracy obtained from the uncleaned (with outliers) and cleaned (without outliers) datasets. As can be seen, all models perform remarkably better when outliers are removed from the dataset, demonstrating the significance of the data-cleaning step in predicting the planetary radius.

\section{Impact of data re-scaling on prediction accuracy}\label{appB}
\begin{figure}
	\includegraphics[width=1.0\columnwidth]{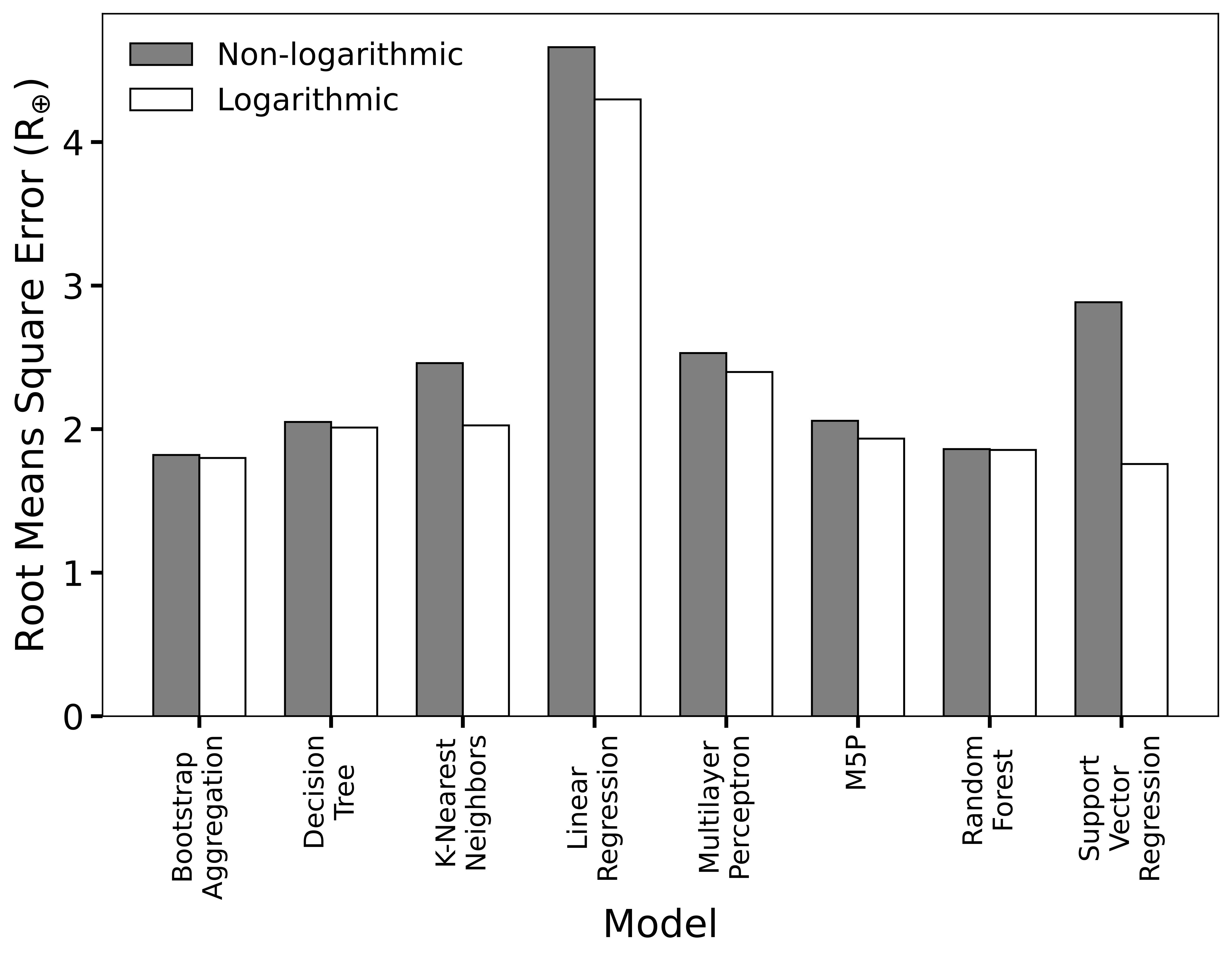}
    \caption{Root means square error (RMSE) values of different ML regression models implemented in logarithmic (white bars) and non-logarithmic (gray bars) datasets. Bootstrap Aggregation and Random Forest models do not show a remarkable difference between logarithmic and non-logarithmic scaling. In contrast, other algorithms, particularly the Support Vector Regression, provide better results on a logarithmic scale.}
    \label{figB1}
\end{figure}

We process both logarithmic and non-logarithmic datasets to understand the effect of data re-scaling on predicting planetary radius. Fig.~\ref{figB1} compares the RMSE values corresponding to different models applied in non-logarithmic and logarithmic datasets. Bootstrap Aggregation and Random Forest slightly differ between logarithmic and non-logarithmic scaling. In contrast, other algorithms, particularly Support Vector Regression, provide better results on a logarithmic scale. Transforming the exoplanet data into a logarithmic space helps handle the wide range of values by compressing them, allowing the ML model better to capture the underlying patterns and relationships within the data. Moreover, logarithm transformation efficiently addresses data skewness and outliers. When data do not follow a normal distribution and contain extreme values, the logarithmic space helps mitigate the influence of outliers by compressing their impact and making the data more symmetrical.

\bsp	
\label{lastpage}
\end{document}